\documentclass[
superscriptaddress,nofootinbib,amsmath,amssymb,prd,twocolumn]{revtex4-2}
\usepackage[colorlinks=true,linkcolor=blue,citecolor=magenta,linktocpage=true]{hyperref}
\usepackage{graphicx}
\usepackage{bm}
\usepackage{hyperref}
\usepackage{mathtools}
\usepackage{soul}
\usepackage{qtree}
\usepackage{booktabs}
\usepackage{tabularx}

\renewcommand{\d}{{\rm d}}

\newcommand{\mrm}[1]{_{\rm #1}}

\makeatletter
\def\l@subsubsection#1#2{}
\makeatother

\begin{document}

\title{Limits on primordial black holes detectability with Isatis: A BlackHawk tool}
\vspace{0.5cm}

\author{Jérémy Auffinger}
\email{j.auffinger@ipnl.in2p3.fr}
\affiliation{Univ Lyon, Univ Claude Bernard Lyon 1,\\ CNRS/IN2P3, IP2I Lyon, UMR 5822, F-69622, Villeurbanne, France}

\begin{abstract}
\vspace{0.5cm}
    Primordial black holes (PBHs) are convenient candidates to explain the elusive dark matter (DM). However, years of constraints from various astronomical observations have constrained their abundance over a wide range of masses, leaving only a narrow window open at $10^{17}\,{\rm g} \lesssim M \lesssim 10^{22}\,$g for all DM in the form of PBHs. We reexamine this disputed window with a critical eye, interrogating the general hypotheses underlying the direct photon constraints. We review 4 levels of assumptions: \textit{i)} instrument characteristics, \textit{ii)} prediction of the (extra)galactic photon flux, \textit{iii)} statistical method of signal-to-data comparison and \textit{iv)} computation of the Hawking radiation rate. Thanks to \texttt{Isatis}, a new tool designed for the public Hawking radiation code \texttt{BlackHawk}, we first revisit the existing and prospective constraints on the PBH abundance and investigate the impact of assumptions \textit{i)-iv)}. We show that the constraints can vary by several orders of magnitude, advocating the necessity of a reduction of the theoretical sources of uncertainties. Second, we consider an ``ideal'' instrument and we demonstrate that the PBH DM scenario can only be constrained by the direct photon Hawking radiation phenomenon below $M\mrm{max} \sim 10^{20}\,$g. The upper part of the mass window should therefore be closed by other means.
\end{abstract}

\maketitle

\tableofcontents

\section{Introduction}

Primordial black holes (PBHs) are early universe objects predicted by lots of inflation models; they could result from large overdensities collapse~\cite{Escriva:2021aeh} as well as more exotic events such as phase transitions or topological defect collapse (see the reviews~\cite{Carr:2020gox,Carr:2020xqk} and references therein or the lecture notes~\cite{Carr:2021bzv}). They are not the outcome of star collapse, hence their mass can span a wide range from the Planck mass $M\mrm{Pl}\simeq 10^{-5}\,$g up to ``stupendously large'' values $M\sim 10^{50}\,$g~\cite{Carr:2020erq}. One most interesting aspect of PBHs is that they can explain all or part of the missing dark matter (DM) density in the universe~\cite{Carr:2020gox,Carr:2020xqk,Carr:2021bzv}). As such, the constraint on their abundance translates into a constraint on the fraction of DM, $f\mrm{PBH} \equiv \Omega\mrm{PBH}/\Omega\mrm{DM}$, they can represent.

Because all BHs evaporate due to Hawking radiation~\cite{Hawking:1974rv,Hawking:1975vcx}, losing their mass in a time related to their initial mass by $\tau \sim M\mrm{init}^3$, their contribution to DM today implies that their initial mass was more than a few times $10^{14}\,$g. Ref.~\cite{Tabasi:2021cxo} underlines that accretion may overcome Hawking radiation during the radiation domination era, causing PBH to grow during this period, so that the mass of PBHs evaporating just now is reduced from $M\lesssim 10^{15}\,$g to $M\sim 10^{14}\,$g. This justifies that we safely show the constraints down to $M \ge M\mrm{min} = 10^{14}\,$g. The emission of Hawking radiation by PBHs, in the form of all kinds of particles and particularly photons, gives rise to observational constraints. PBHs with mass $M \sim 10^{20}\,$g emit in the keV band; as the mean energy of Hawking radiation is inversely proportional to the PBH mass $\langle E \rangle \propto 1/M$, heavier PBHs $M \ge M\mrm{max} = 10^{20}\,$g are very hard to see through their photon Hawking radiation, as will be shown in the following.

During the last years, numerous constraints have been established on the PBH abundance in this disputed mass range $[M\mrm{min},M\mrm{max}]$, with at stake the answer to the question of whether light PBHs are indeed the elusive DM. What we will henceforth call \textit{direct} photon constraints are set by comparing the PBH emission spectrum to the galactic center (GC), or any other target, photon flux~\cite{Laha:2020ivk,Coogan:2020tuf,Coogan:2021rez}, to the stacked isotropic X-ray or gamma-ray backgrounds (denoted as EXGB)~\cite{Carr:2009jm,Arbey:2019vqx,Ballesteros:2019exr,Cai:2021zxo,Mittal:2021dpe} or to a combination of the two~\cite{Ray:2021mxu,Iguaz:2021irx,Ghosh:2021gfa}. There also exist \textit{direct} photon (and neutrino) constraints on the local rate of PBH final burst as they reach the end of their evaporation, but this is specific to a fine-tuned PBH mass and thus we do not consider them here (see the very complete article~\cite{Capanema:2021hnm}). \textit{Indirect} photon constraints are related e.g.~to the change in the re-ionization history at the cosmic microwave background~\cite{Cang:2020aoo,Acharya:2020jbv}. Other studies are based on the direct~\cite{Boudaud:2018hqb} or indirect detection of electrons-positrons~\cite{Laha:2019ssq,Dasgupta:2019cae,Acharya:2020jbv,Dutta:2020lqc,Kim:2020ngi,Laha:2020vhg,Chan:2020zry,Mittal:2021egv,Keith:2021guq,Cang:2021owu,Mukhopadhyay:2021puu,Lee:2021qhe,Lu:2021rnz,Siegert:2021upf,Natwariya:2021xki,Saha:2021pqf}; and finally there are some papers that focus on neutrino constraints~\cite{Dasgupta:2019cae,Wang:2020uvi,Calabrese:2021zfq,DeRomeri:2021xgy}. Most of the more recent studies now take into account the effect of having a non-zero PBH spin, which enhances the photon emission and thus results in more stringent constraints for Kerr (rotating) PBHs~\cite{Arbey:2019vqx,Dasgupta:2019cae,Laha:2020ivk,Laha:2020vhg,Iguaz:2021irx,DeRomeri:2021xgy,Natwariya:2021xki,Cang:2021owu,Ghosh:2021gfa,Saha:2021pqf}, or an extended PBH mass distribution, that should be more realistic regarding the channels of PBH formation~\cite{Arbey:2019vqx,Dasgupta:2019cae,Laha:2019ssq,Chan:2020zry,Iguaz:2021irx,Cai:2021zxo,DeRomeri:2021xgy,Cang:2021owu,Mittal:2021dpe}. Exotic studies have also derived constraints on non-standard (other than Kerr) PBHs~\cite{Johnson:2020tiw,Arbey:2021yke,Kritos:2021nsf} or PBH-DM mixed scenarios~\cite{Carr:2020mqm,Calabrese:2021src,Chao:2021orr,Kadota:2021jhg}. To do so, recent constraints make use of the public code \texttt{BlackHawk}~\cite{Arbey:2019mbc,Arbey:2021mbl}\footnote{\texttt{BlackHawk} is available at \url{https://blackhawk.hepforge.org}} to compute the precise Hawking radiation spectra, and we are using the latest version \texttt{2.1} of this code in our analysis. Photon constraints are handy thanks to several simplifications: photons are massless particles, thus any PBH mass is compatible with the emission of photons with the corresponding mean energy; photons travel in straight lines, so their detection does not rely on complex propagation models; and the detection of photons by all kinds of instruments is a mastered technique that does not suffer from large measure uncertainties and has already been performed accurately in a very wide range of energies. This is why we focus on photons in this study, leaving electrons (all simplifications do not hold) and neutrinos (only the last simplification does not hold) for future work. We further restrict the scope of our analysis to most minimalist Schwarzschild PBHs with a monochromatic mass distribution and unclustered spatial distribution. A discussion on more elaborate PBH models is postponed to the last Section of this paper.

As stated above, the fraction $f\mrm{PBH}$ is already partially constrained in the $M\mrm{min}-10^{16}\,$g range, leaving small doubt on the possibility that PBHs can constitute all of DM in the low mass part of this window, with constraints robustly set to $f\mrm{PBH} < 10^{-5}$. However, the high mass range $10^{16}-10^{22}\,$g has been the subject of multiple studies that claimed exclusion of $100\%$ PBH-DM up to $10^{17}$ and even $10^{18}\,$g (e.g.~EXGB constraints in the case of an extended mass distribution of spinning PBHs~\cite{Arbey:2019vqx}). The aim of this article is to make quantitatively explicit the underlying assumptions and approximations, to contextualize the claimed limit validity and to underline the way they can (or cannot) be compared to each other. This is of utmost importance because not the whole mass window that is currently scrutinized for $100\%$ PBH-DM is accessible to Hawking radiation constraints, as will be demonstrated; extended mass functions that are fitted to the available remaining window constrain back the fine-tuned PBH formation models (and thus the early universe conditions) used to derive them; and the robustness estimation of the limits can help design optical instruments that would be sensitive to the most extreme parameter space accessible to Hawking radiation measures. We regroup the assumptions into 4 categories used throughout this paper, 
all of which discussed in details in the following:
\begin{itemize}
    \item[\textit{i)}] instrument characteristics,
    \item[\textit{ii)}] computation of the (extra)galactic photon fluxes,
    \item[\textit{iii)}] statistical treatment,
    \item[\textit{iv)}] computation of the Hawking radiation.
\end{itemize}

This paper is organized as follows: Section~\ref{sec:hawking_rad} gives the required basic formulas for Hawking radiation; Section~\ref{sec:nomenclature} proposes a nomenclature of type \textit{iii)} assumptions; Section~\ref{sec:results} discusses the robustness of the current existing and prospective Hawking radiation constraints on PBHs and Section~\ref{sec:engineering} determines the characteristics and limits of an instrument for high PBH mass Hawking radiation studies; we conclude in Section~\ref{sec:conclusion}. Throughout the paper, we use natural units $c = G = k\mrm{B} = \hbar = 1$, making the constants appear uniquely when dimensionality is unclear.

\section{Hawking radiation constraints: principle}
\label{sec:hawking_rad}

\subsection{Basic formulas}

Since it has been shown that BHs emit quasi-thermally all particles in the ``spectrum of Nature''~\cite{Hawking:1974rv,Hawking:1975vcx}, people have tried to extract a PBH signal from the observational data, without success so far. This absence of signal has been interpreted as a constraint on the PBH abundance over a wide range of masses. For an initial mass from the Planck scale $M\mrm{Pl} \approx 10^{-5}\,$g to some $M\mrm{min} = 10^{14}\,$g, PBHs would have already evaporated away by now, forbidding them to be a sizeable component of the DM today, but their Hawking radiation could have left imprints in the big bang nucleosynthesis element abundances or in the cosmic microwave background~\cite{Carr:2020gox}. Above an initial mass of some $M\mrm{min}$, PBHs would still be around, filling the universe with all kind of radiation, from photons to hadronized jets.

The emission rate of a massless particle $i$ with spin $s_i$ by a Schwarzschild BH is given by~\cite{Hawking:1975vcx}
\begin{equation}
    Q_i(E,t) \equiv \dfrac{\d^2 N_i}{\d t\d E} = \dfrac{1}{2\pi} \dfrac{\Gamma_i}{e^{E/T} - (-1)^{s_i}}\,,\label{eq:emission_rate}
\end{equation}
where $\Gamma_i$ is the ``greybody factor'' (GF) that encodes the probability that the emitted particle reaches spatial infinity away from the BH horizon. The GFs deviate from the pure blackbody spectrum, with a power-law suppression at lower energies that depends on the particle's spin. Those GFs were numerically computed in the public code \texttt{BlackHawk} that we use in this study.\footnote{For more information about the numerical methods please refer to the \texttt{BlackHawk} manual~\cite{Arbey:2019mbc}.} The BH horizon temperature is related to the BH mass through
\begin{equation}
    T = \dfrac{1}{8\pi M}\,,\label{eq:temperature}
\end{equation}
so that when the mass decreases because of evaporation, the temperature increases. The emission rate of Eq.~\eqref{eq:emission_rate} is further cutoff at the particle rest mass $E > \mu_i$, which is trivial for massless photons. It should be noted that both the BH temperature $T$ and the GFs $\Gamma_i$ depend strongly on the BH metric (see e.g.~\cite{Arbey:2021jif,Arbey:2021yke} for the general formulas for spherically symmetric BH solutions).

Once the instantaneous emission rates $Q_i$ of all particles $i$ are known, they can be integrated over to obtain the BH mass loss rate~\cite{Page:1976df}
\begin{equation}
    \mathcal{F}(M) \equiv M^2 \int_0^{+\infty} E \sum_i Q_i(E)\,\d E\,,
\end{equation}
which determines the BH mass evolution equation
\begin{equation}
    \dfrac{\d M}{\d t} = -\dfrac{\mathcal{F}(M)}{M^2}\,.
\end{equation}
Estimates of the function $\mathcal{f}(M)$ can be found e.g.~in~\cite{Cheek:2021odj}. Inside \texttt{BlackHawk} we use a full numerical result. Considering that $\mathcal{f}(M)$ is a constant, we obtain a relationship between the BH initial mass $M\mrm{init}$ and its lifetime $\tau$
\begin{equation}
    \tau(M\mrm{init}) \sim 13.9\times 10^{9}\,{\rm yr} \left(\dfrac{M\mrm{init}}{5\times 10^{14}\,{\rm g}}\right)^3.
\end{equation}
A more detailed study of the evolution of BHs can be found in e.g.~\cite{Arbey:2019jmj}.

We would like to emphasize that the above calculations assume the Hawking formula~\eqref{eq:emission_rate}~\cite{Hawking:1974rv,Hawking:1975vcx} and the GF computation inside \texttt{BlackHawk}, based upon semi-classical general relativity, hold throughout (most) of the BH history. They may break down only at the very end of evaporation $t\sim \tau$, where $\tau$ is the usual BH lifetime calculated above, when $M\sim M\mrm{Pl}$ and quantum gravity effects become relevant. This has no effect on the constraints under discussion. However, it has been claimed that the so-called ``memory burden'' of BHs~\cite{Dvali:2018xpy}---their capacity to store information---can slow down the evaporation process to extremely low rates or even stop it, within timescales $t<\tau$. If that were true, the evaporation constraints set by e.g.~photon background measurement would be dramatically alleviated and PBHs of mass $M\ll 10^{14}\,$g could represent a fraction or all of DM~\cite{Dvali:2020wft}. Hereafter, we assume that the usual paradigm holds until $M\sim M\mrm{Pl}$.

This quasi-thermal emission, that we call ``primary'', is not the final output of Hawking radiation. Most particles in the Standard Model are not stable (weak gauge bosons, charged leptons) or cannot exist outside confined hadrons (gluons and quarks). The primary emission must be convolved with analytical or numerical branching ratios ${\rm Br}_{i\rightarrow j}$ to obtain the ``secondary'' spectra of stable particles that can be detected in instruments (photons, electrons, neutrinos, protons and their antiparticles). The secondary spectrum for particle $j$ is then
\begin{equation}
    \tilde{Q}_j(E,t) \equiv \dfrac{\d^2 N_j}{\d t\d E} = \sum_i \int_0^{+\infty}\!\!{\rm Br}_{i\rightarrow j}(E,E^\prime)\, Q_i(E^\prime,t)\,\d E^\prime\,.\label{eq:secondary}
\end{equation}
In particular, $\tilde{Q}_\gamma$ is the rate of emission of \textit{direct} photons from PBHs, meaning that no interaction with the interstellar medium (ISM) or with other astronomical objects has been considered (scattering, absorption...). Inside \texttt{BlackHawk v2.1} we use the \texttt{Python} package \texttt{Hazma}~\cite{Coogan:2019qpu}, that is relevant for low energy hadronization and decays ($E \lesssim$ some GeV), to compute the branching ratios ${\rm Br}_{i\rightarrow j}$.\footnote{The branching ratio from photon to photon ${\rm Br}_{\gamma\rightarrow \gamma}(E,E^\prime)$ is a Dirac function $\delta(E-E^\prime)$.} This package considers that pions are emitted as fundamental particles instead of single quarks and gluons, and subsequently decayed into photons and leptons. \texttt{Hazma}, which relies on analytical decay and final state radiation formulas, undoubtedly suffers from inherent approximations. We should also implement Bremsstrahlung effects for emitted charged particles~\cite{Page:2007yr} that may dominate at very low energy, around the keV scale. Using \texttt{Hazma}, Ref.~\cite{Coogan:2020tuf} showed that as the emission of all the massive particles (the lightest being the electron) is exponentially suppressed for PBHs with mass $M\gtrsim 10^{17}\,$g only primary photons, neutrinos and maybe gravitons are radiated. Secondary photons however dominate the spectrum for PBHs with lower masses, with a mean energy $\langle \tilde{E} \rangle$ well below the PBH temperature. The implementation of \texttt{Hazma} inside \texttt{BlackHawk v2.0} was a necessary improvement compared to the extrapolation of the \texttt{PYTHIA}~\cite{Sjostrand:2014zea} and \texttt{HERWIG}~\cite{Bellm:2015jjp} results used in the previous version \texttt{BlackHawk v1.2}. The PBH nature (beyond Schwarzschild) and mass distribution, as well as the computation of the GFs and branching ratios, belong to the type \textit{iv)} assumptions.

In the following two subsections, we give the basic formulas for (extra)galactic PBH photon flux computation, together with a discussion of related type \textit{ii)} assumptions.  Both contributions are important because coincidentally, they are of the same order of magnitude at MeV energies for PBHs right in the relevant Hawking radiation window $M\sim 10^{15}-10^{18}\,$g (a fact that was first noted by Ref.~\cite{Wright:1995bi}). Low energy photons principally come from the redshifted extragalactic spectrum while high energy photons originate from the galaxy~\cite{Ray:2021mxu,Iguaz:2021irx,Ghosh:2021gfa}.

\subsection{Galactic flux}

If PBHs represent some fraction of DM, then they are present in the Milky way halo with some spatial distribution $\rho\mrm{gal}(r)$ (in galactocentric coordinates) which we take as spherical for simplicity. The photons they emit through Hawking radiation propagate in straight lines from their origin PBH to Earth. Thus, the flux of GC PBH photons received by an instrument on Earth per unit energy, time, surface and solid angle is\footnote{These formulas are given in some form in e.g.~\cite{Laha:2020ivk,Coogan:2021rez,Coogan:2020tuf,Ray:2021mxu,Iguaz:2021irx,Ghosh:2021gfa}.}
\begin{equation}
    \dfrac{\d \Phi_\gamma^{\rm gal}}{\d E} = \dfrac{1}{A\mrm{gal}}\dfrac{J\mrm{gal}}{4\pi} \tilde{Q}_\gamma(E)\,,\label{eq:flux_gal}
\end{equation}
where
\begin{equation}
    J\mrm{gal} \equiv \dfrac{1}{\Delta\Omega} \int_{\Delta\Omega} \d\Omega \int_{\rm LOS} \rho\mrm{gal}(r(l,\Omega))\,\d l\,,
\end{equation}
with $\Delta\Omega$ the field of view considered and $A\mrm{gal} \equiv M$ is the normalisation constant for a monochromatic mass $M$ distribution of Schwarzschild PBHs. Note that this formula could be adapted for any compact source other than the GC with the relevant surface factor $J$ integrated over the volume of that source. We see that the flux depends on the mass distribution of DM in the Milky Way halo and on the precise location $R_0$ of the Solar System in this halo, contained in the definition of the line of sight (LOS)~\cite{Ghosh:2021gfa}.

\subsection{Extragalactic flux}

PBHs form in the early universe, with a formation time $t\mrm{form}$ related to their initial mass by~\cite{Carr:2020gox}
\begin{equation}
    t\mrm{form} \sim 10^{-23}\,{\rm s}\left( \dfrac{M\mrm{init}}{10^{15}\,{\rm g}} \right),
\end{equation}
assuming radiation domination. Hence, they emit particles through Hawking radiation during all cosmological eras until today, and this continuous flux piles up with a dilution factor $a(t) = 1 + z(t)$. The flux on an instrument today per unit energy, time, surface and solid angle is given by\footnote{These formulas can be found in some form in~\cite{Arbey:2019vqx,Ballesteros:2019exr,Cai:2021zxo,Mittal:2021dpe,Ray:2021mxu,Iguaz:2021irx,Ghosh:2021gfa}.}
\begin{equation}
    \dfrac{\d\Phi_\gamma^{\rm egal}}{\d E} = \dfrac{1}{A\mrm{egal}}\dfrac{c}{4\pi}\int_{t\mrm{min}}^{t\mrm{max}}\hspace{-0.5cm} \d t \int_0^{+\infty}\hspace{-0.5cm} a(t)\,\tilde{Q}_\gamma(t,E = a(t)E^\prime)\,\d E^\prime\,,\label{eq:flux_egal}
\end{equation}
with $t\mrm{min} = t\mrm{CMB}$ for photons, because the universe is opaque to light before that, and $t\mrm{max} = \min(\tau,t\mrm{today})$. The normalization constant for a monochromatic mass $M$ distribution of Schwarzschild PBHs is $A\mrm{egal} \equiv M/\rho\mrm{DM}$ where $\rho\mrm{DM} \equiv \Omega\mrm{DM}\rho\mrm{crit}$ is the global density of DM today. We see that the flux depends on the value of the cosmological DM density fraction, but also on the redshift history of the universe; we have further neglected the optical depth of the ISM (line absorption of light) apart from the cut-off at the CMB time in integral~\eqref{eq:flux_egal}.

\section{A nomenclature of constraints}
\label{sec:nomenclature}

In the previous Section, we have computed the flux of photons emitted directly by PBHs, with a galactic and an extragalactic contribution. Now, we detail how the fluxes \eqref{eq:flux_gal} and \eqref{eq:flux_egal} are compared to different sets of data to derive PBH abundance constraints. This discussion is quite trivial and cumbersome but we think that it is of central importance to recall the basics. Altogether, this Section thus deals with type \textit{iii)} assumptions.

\subsection{Existing data}

The first and very classical method to obtain a constraint on the PBH abundance from Hawking radiation consists of directly comparing the predicted flux for some PBH mass and abundance --- we recall that we focus only on a monochromatic distribution of Schwarzschild PBHs --- and some set of data measured by an instrument. Photon data are either presented in a differential (energy/events per unit time) or integrated (total energy/events) form, as a function of photon energy. Both should be equivalent as we do not expect the PBH signal to vary during the time of observation. Then, to constrain a signal, one can compare the spectrum to each energy bin of the data set, asking that the PBH photon flux does not overrun the measured flux. One can also consider the whole instrument energy band and compare the integrated quantities. This is a model independent method, but there are already two ways to compare the signal to the data, and there remains a choice to make as for the confidence level (CL) we chose to exclude PBHs (data, data $+\sigma$, data $+2\sigma$, ...). One can also fit the data with some function (from a simple segmented function to a complex analytical formula), motivated by a physical interpretation or not. This is already a model dependent approximation, as it infers data in unmeasured energy bins from data in measured ones. One can finally go one step further, by assuming that some fraction of the measured signal comes from astronomical sources. Thus, there is even less available parameter space for the PBH abundance. This is highly model dependent, and contains a hidden feedback loop: most astronomical backgrounds are estimated (calibrated) thanks to the data. Furthermore, the data are often ``cleaned'' using catalogs of identified point-sources to obtain the diffuse components. This introduces a bias that we have used in the present analysis: if PBHs are highly clustered in the Galaxy, they would resemble point-like sources and be cleaned away in the diffuse component search procedure. Constraints for clustered PBHs require a dedicated treatment (see e.g.~\cite{Belotsky:2014kca,Belotsky:2018wph}).

\subsection{Prospective instruments}

When data is not yet available in some energy range, one can put a conservative constraint on the PBH signal by saying that if the prospective instrument designed to explore this particular energy range is built and measures nothing, then this means that the signal is below the sensitivity of the instrument, with some confidence level. When data is already available, it is very complicated to decide what to do. In fact, the more precise instrument to come can totally revolutionize the measures due to an error of appreciation of the functioning of the previous ones. An independent conservative constraint would be to predict what sensitivity to a PBH signal some instrument can \textit{in principle} reach, assuming that all the signal comes from PBHs or that some (model-dependent) background is to subtract beforehand. This is quite radical and nobody expects that all the measures taken by long-time working instruments are to be thrown away. A more reasonable method is to build a model of background, and to use it to estimate what would be the prominence of the PBH signal ``above'' the background (signal-to-noise ratio --- SNR --- method or $\chi^2$ method). This also contains a feedback loop: background models are often calibrated thanks to the older instrument data, thus a deviation from the expected background compensated by a PBH signal would appear as no signal at all. Once again, the PBH signal and the background can be compared over the whole energy range or inside each energy bin separately.

\subsection{Summary}

\begin{figure}
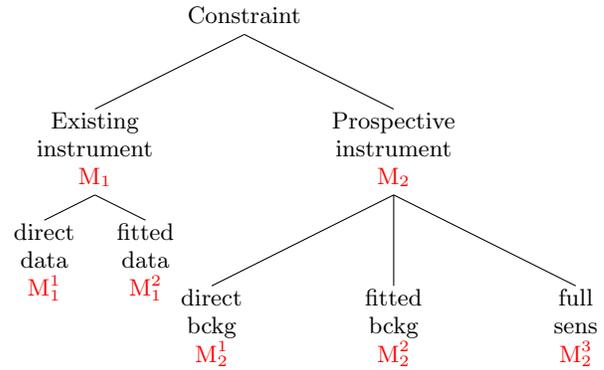

    \hspace{-0.5cm}
    \Tree[.Constraint [.{Existing\\instrument\\\textcolor{red}{M$_1$}} !\qsetw{0.1\textwidth} {direct\\data\\\textcolor{red}{M$_1^1$}} !\qsetw{0.1\textwidth} {fitted\\data\\\textcolor{red}{M$_1^2$}} ] [.{Prospective\\instrument\\\textcolor{red}{M$_2$}} !\qsetw{0.12\textwidth} {direct\\bckg\\\textcolor{red}{M$_2^1$}} !\qsetw{0.15\textwidth} {fitted\\bckg\\\textcolor{red}{M$_2^2$}} !\qsetw{0.12\textwidth} {full\\sens\\\textcolor{red}{M$_2^3$}} ] ]
    \caption{The different constraint methods. We have restricted the tree to methods we have encountered in the literature: for existing instruments, ``direct data'' means that the instrument data points are used, and ``fitted data'' that a fit running through the data points is chosen; for prospective instruments, ``direct bckg'' means that the data points used as agnostic background, ``fitted bckg'' that some fit is used and ``full sens'' that pre-existing data are ignored and the new detector maximal sensitivity is used. The designation we use in the text are: ``type 1'' methods M$_1$ with sub-methods M$_1^1$ and M$_1^2$ for existing instruments; ``type 2'' methods M$_2$ with sub-methods M$_2^1$, M$_2^2$ and M$_2^3$ for prospective instruments.}
    \label{fig:tree}
\end{figure}

Hence, we see that even the simplest data-signal comparison method contains non-trivial features that complicate the comparison between constraints set with different choices. This is even worse for the prospective instrument methods as the dependency to the background modelling is complex. We do not intend to build a hierarchy of the methods chosen by different authors but just to quantitatively highlight their differences. Ref.~\cite{Coogan:2020tuf} is the first, to our knowledge, to compare the PBH constraints from different prospective instruments with rigorously the same statistical method. One must also check that instrument characteristics and theoretical choices for the PBH flux calculation are coherent from one study to another. We have built a tree in Fig.~\ref{fig:tree} to summarize the methods described above, restricted to the ones used in the literature cited in the Introduction. \textit{Please refer to this tree for all the subsequent abbreviated method nomenclature.}

\section{A numerical tool: the Isatis program}
\label{sec:results}

For the sake of this study, we have designed \texttt{Isatis}, a numerical public tool that relies on the \texttt{BlackHawk} PBH spectra to compute the constraints with a controlled set of assumptions. The code is presented in details in the Appendix~\ref{app:Isatis}. The idea is to use \texttt{BlackHawk} to obtain the \textit{direct} PBH photon spectrum, and then to derive the constraint on the PBH abundance for a list of optical instruments, existing or prospective. All constraint assumption types \textit{i)-iv)} listed in the Introduction, with different statistical methods from the nomenclature given in Fig.~\ref{fig:tree} available, can be modified. Hence, the quantitative individual impact of the assumptions on the PBH ``direct'' photon constraints can be investigated. This allows to compare consistently the constraints from the literature. Identifying the dominant parameters further gives an insight on where to look for improvements in the constraints. In this Section, we revisit the literature constraints from GC and EXGB PBH photon fluxes and examine type \textit{iii)} assumption impact.

\subsection{Existing instruments}
\label{subsec:existing}

\begin{figure*}
    \centering
    \includegraphics[scale = 1]{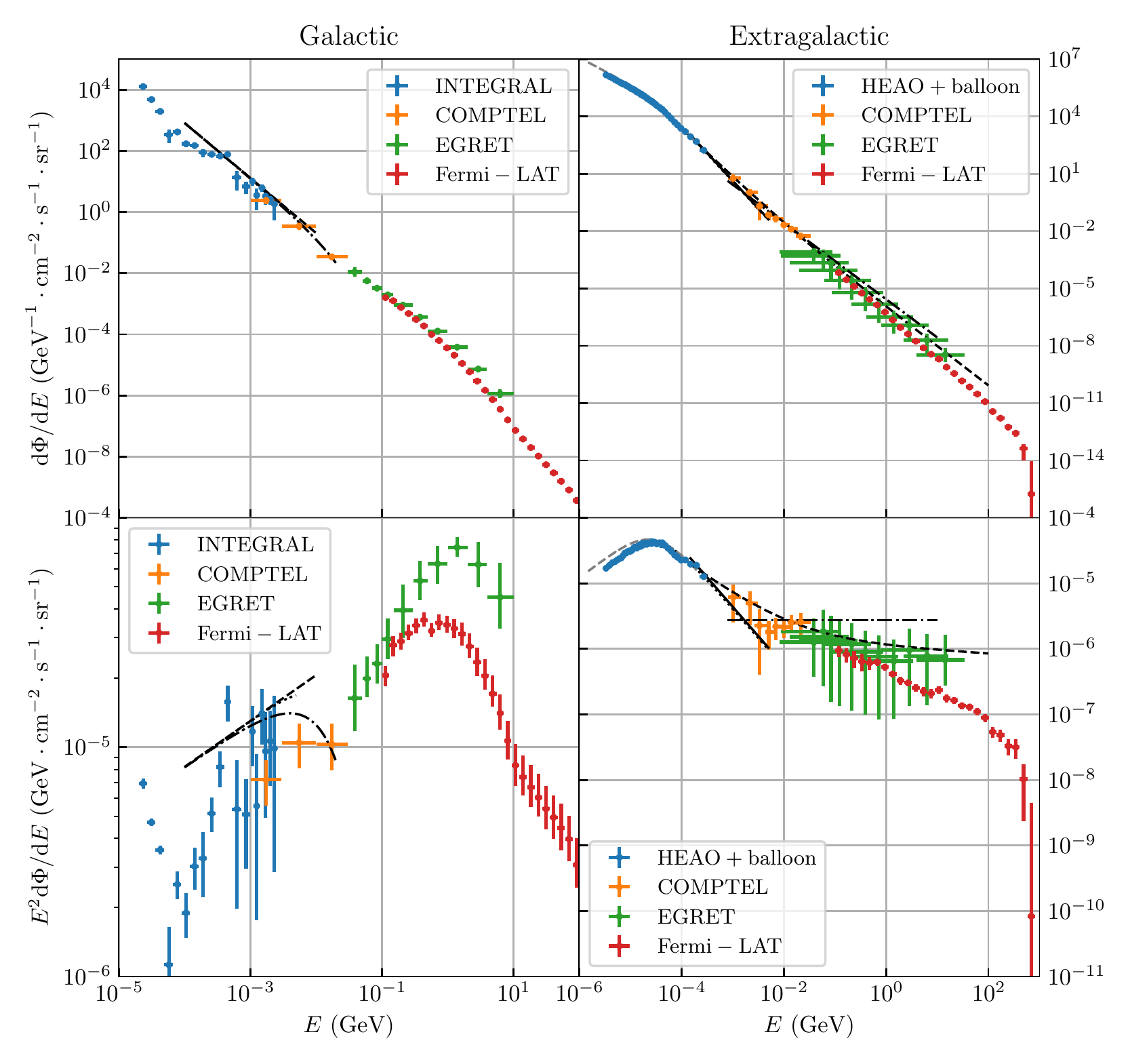}
    \caption{\textbf{Left:} GC photon flux measured by INTEGRAL~\cite{Bouchet:2011fn} ($l\in[-30,+30]$, $b\in[-15,+15]$), COMPTEL~\cite{Strong1994,Bouchet:2011fn} ($l\in[-30,+30]$, $b\in[-15,+15]$), EGRET~\cite{Strong1996,Strong:1998ck} ($l\in[-30,+30]$, $b\in[-5,+5]$) and Fermi-LAT~\cite{Strong:2011pa} ($l\in[-30,+30]$, $b\in[-10,+10]$). For completeness, we show the following background models: \cite{Beacom:2005qv} (dashed), \cite{Bartels:2017dpb} (dot-dashed) and \cite{Ray:2021mxu} (dotted).
    \textbf{Right:} Isotropic extragalactic photon flux measured by HEAO+balloon~\cite{Gruber:1999yr}, COMPTEL~\cite{Ruiz-Lapuente:2015yua}, EGRET~\cite{Strong:2004ry} and Fermi-LAT~\cite{Fermi-LAT:2014ryh} (model A). For completeness, we show the following background models: \cite{Gruber:1999yr} (dashed, black), \cite{Moretti:2008hs} (dashed, grey), \cite{Boddy:2015efa} (dot-dashed), \cite{Ballesteros:2019exr} (dotted),   \cite{Ray:2021mxu} (solid).
    \textbf{Upper} and \textbf{lower} panels show $\d\Phi/\d E$ and $E^2\d\Phi/\d E$ as both representations give useful insight.\label{fig:backgrounds}}
\end{figure*}

\begin{figure*}
    \centering
    \includegraphics[scale = 1]{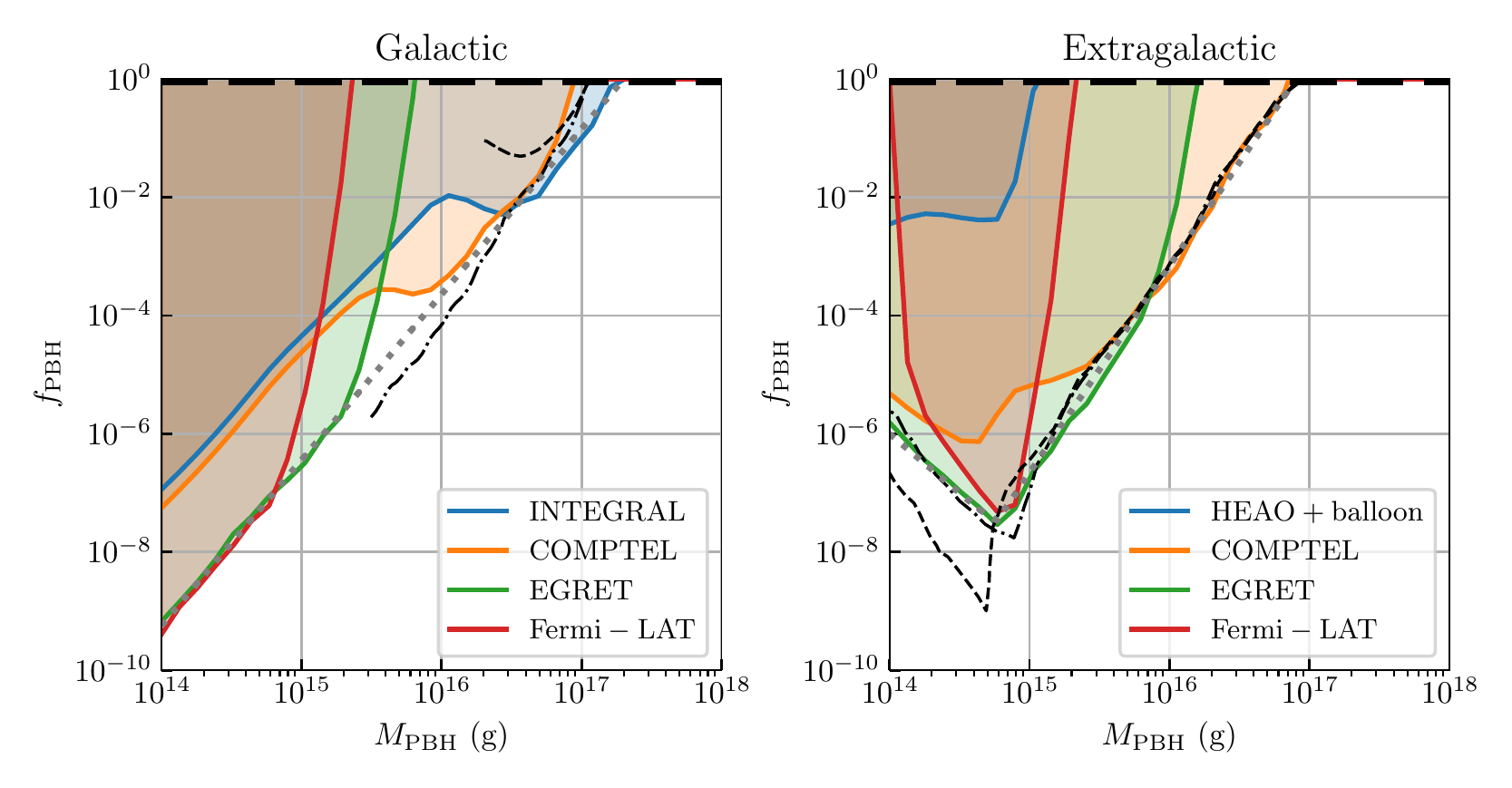}
    \caption{\textbf{Left:} PBH constraints from GC observations of the 4 instruments of Fig.~\ref{fig:backgrounds}. For comparison, we show the limits set by Ref.~\cite{Laha:2020ivk} for INTEGRAL (dashed) and Ref.~\cite{Coogan:2020tuf} for COMPTEL (dot-dashed), stressing the fact that the statistical method is not the same as here.
    \textbf{Right:} PBH constraints from EXGB observations of the 4 instruments of Fig.~\ref{fig:backgrounds}. For comparison, we show the limits set by Ref.~\cite{Carr:2009jm} (dashed) and Ref.~\cite{Arbey:2019vqx} (dot-dashed).
    \textbf{Method:} Limits from this work (shaded areas) are computed with method M$_1^1$ using the central values of the Fig.~\ref{fig:backgrounds} fluxes (${\rm CL} = 0$ in Eq.~\eqref{eq:existing}).
    }
    \label{fig:existing}
\end{figure*}

In Fig.~\ref{fig:backgrounds} (left panels), we show the photon flux measured by 4 instruments in the GC direction (latitude $b$ and longitude $l$ close to 0 in galactocentric coordinates), spanning the $10\,{\rm keV}-100\,$GeV energy range: INTEGRAL~\cite{Bouchet:2011fn}, COMPTEL~\cite{Strong1994,Bouchet:2011fn}, EGRET~\cite{Strong1996,Strong:1998ck} and Fermi-LAT~\cite{Strong:2011pa}.\footnote{Be careful that the field of view around the GC on which the data as been averaged is not the same for each instrument, causing probable normalization incompatibilities on that plot.} In the same figure (right panels), we show the EXGB measured by 4 instruments, spanning the $1\,{\rm keV}-1\,$TeV energy range: HEAO+balloon~\cite{Gruber:1999yr},\footnote{The X-ray sky has since been explored by several more recent instruments, as reviewed e.g.~in~\cite{Brandt:2021lch}.} COMPTEL~\cite{Ruiz-Lapuente:2015yua}, EGRET~\cite{Strong:2004ry} and Fermi-LAT~\cite{Fermi-LAT:2014ryh} (model A). We also show various fitting models (see legend), quite accurate in some energy ranges but unsuccessful in reproducing the whole set of data.

To obtain the PBH constraints, we must use method M$_1$. We show the results for the submethod M$_1^1$ in Fig.~\ref{fig:existing}: direct comparison of the signal to the data requiring that
\begin{align}
    f\mrm{PBH}\int_{E\mrm{low}^{\rm X}}^{E\mrm{up}^{\rm X}} \dfrac{\d \Phi^{\rm PBH}\mrm{gal/egal}}{\d E}\,\d E \le\, &(E\mrm{up}^{\rm X} - E\mrm{low}^{\rm X})\,\times\label{eq:existing} \\ &\left[\dfrac{\d \Phi\mrm{gal/egal}^{\rm X}}{\d E} + {\rm CL}\times\Delta\Phi^{\rm X} \right],\nonumber
\end{align}
in each energy bin, where the fluxes $\d\Phi\mrm{gal/egal}/\d E$ are given by Eqs.~\eqref{eq:flux_gal} and \eqref{eq:flux_egal}, $E\mrm{low/up}^{\rm X}$ are the bounds of the energy bins of instrument X and $\d\Phi\mrm{gal/egal}^{\rm X}/\d E$ (resp.~$\Delta\Phi^{\rm X}$) is the photon flux (resp.~error bar) measured by X and shown in Fig.~\ref{fig:backgrounds}. CL is the confidence level, translated in the number of error bars considered above or below the central value of the data points. We have chosen a NFW DM profile~\cite{Navarro:1996gj} with the ``convenient'' set of parameters from~\cite{McMillan:2011wd} for the galactic flux, and the standard radiation domination from CMB ($t \approx 1.2\times 10^{13}\,$s) to today ($t \approx 4.4\times 10^{17}\,$s) for the extragalactic flux, with $\Omega\mrm{DM}$ given by Planck~\cite{Planck:2018vyg}. In Fig.~\ref{fig:existing}, we have performed an empirical power-law fit to the most stringent constraints, with parameters
\begin{equation}
    f\mrm{PBH} \sim (M\,{\rm (g)}/2\times 10^{17})^{2.8}\,,
\end{equation}
for the GC, which overestimates the results at $\sim5\times 10^{15}\,$g and
\begin{equation}
    \left\{\begin{array}{ll}
        f\mrm{PBH} \sim (M\,{\rm (g)}/10^{11})^{-2}\,, & M \lesssim 8\times 10^{14}\,{\rm g} \\
        f\mrm{PBH} \sim (M\,{\rm (g)}/8\times 10^{16})^{3.5}\,, & M \gtrsim 8\times 10^{14}\,{\rm g} 
    \end{array}\right.
\end{equation}
for the EXGB, which can be compared to~\cite{Carr:2020xqk}.

As an example, let us now examine the impact of the CL parameter, which takes into account the error bars of the photon fluxes. This is a type \textit{iii)} assumption. To do so, we repeat the analysis of Fig.~\ref{fig:existing} with ${\rm CL} = \{-1,1,2\}$ (lower error bar, upper error bar and twice the upper error bar), and present in Table~\ref{tab:CL_impact} the maximum relative discrepancy in $f\mrm{PBH}$, as compared to the case ${\rm CL} = 0$ (central values). Obviously, instruments with large error bars are the most affected: looking at Eq.~\eqref{eq:existing}, we easily deduce that an increase of the instrument flux by a factor $\alpha$ results in a constraint relieved by the same factor $\alpha$. This is directly shown by the linearity between CL and the relative increase in $f\mrm{PBH}$ in Table~\ref{tab:CL_impact} (slightly broken for asymmetric error bars). We observe that the CL parameter can have an impact of up to a factor of a few on $f\mrm{PBH}$ for instruments with very large error bars.

The impact would be exactly of the same kind if the flux data are replaced by some fitting function: the relative discrepancy in the constraint would be directly proportional to the relative discrepancy between the data and the fit. On the other hand, an increase of the PBH signal by a factor $\alpha$ results in a constraint more stringent by the same factor $\alpha$.

\begin{table}[t]
    \centering
    \begin{tabular*}{\columnwidth}{l@{\extracolsep{\fill}}c@{\extracolsep{\fill}}c@{\extracolsep{\fill}}c@{\extracolsep{\fill}}c}
        \toprule
        instrument & $-1$ & $0$ & $+1$ & $+2$ \\
        \midrule
        galactic $\left\{ \begin{array}{l}
            {\rm HEAO+balloon} \\
            {\rm COMPTEL} \\
            {\rm EGRET} \\
            {\rm Fermi-LAT}
        \end{array} \right.$ & $\begin{array}{c}
            -0.68 \\
            -0.21 \\
            -0.40 \\
            -0.10
        \end{array}$ & $\begin{array}{c}
            0 \\
            0 \\
            0 \\
            0
        \end{array}$ & $\begin{array}{c}
            +0.68 \\
            +0.21 \\
            +0.40 \\
            +0.10
        \end{array}$ & $\begin{array}{c}
            +1.36 \\
            +0.44 \\
            +0.81 \\
            +0.20 
        \end{array}$\\
        \midrule
        extragalactic $\left\{ \begin{array}{l}
            {\rm INTEGRAL} \\
            {\rm COMPTEL} \\
            {\rm EGRET} \\
            {\rm Fermi-LAT}
        \end{array} \right.$ & $\begin{array}{c}
            -0.08 \\
            -0.80 \\
            -0.70 \\
            -0.24
        \end{array}$ & $\begin{array}{c}
            0 \\
            0 \\
            0 \\
            0
        \end{array}$ & $\begin{array}{c}
            +0.05 \\
            +0.80 \\
            +1.30 \\
            +0.20
        \end{array}$ & $\begin{array}{c}
            +0.11 \\
            +1.61 \\
            +2.59 \\
            +0.40
        \end{array}$ \\
        \bottomrule
    \end{tabular*}
    \caption{Relative maximum increase in $f\mrm{PBH}$ for ${\rm CL} = \{-1,0,1,2\}$ relative to the case ${\rm CL} = 0$ for the instruments of Fig.~\ref{fig:existing}.}
    \label{tab:CL_impact}
\end{table}

\subsection{Prospective instruments}
\label{subsec:prospective}

\begin{figure*}
    \centering
    \includegraphics[scale = 1]{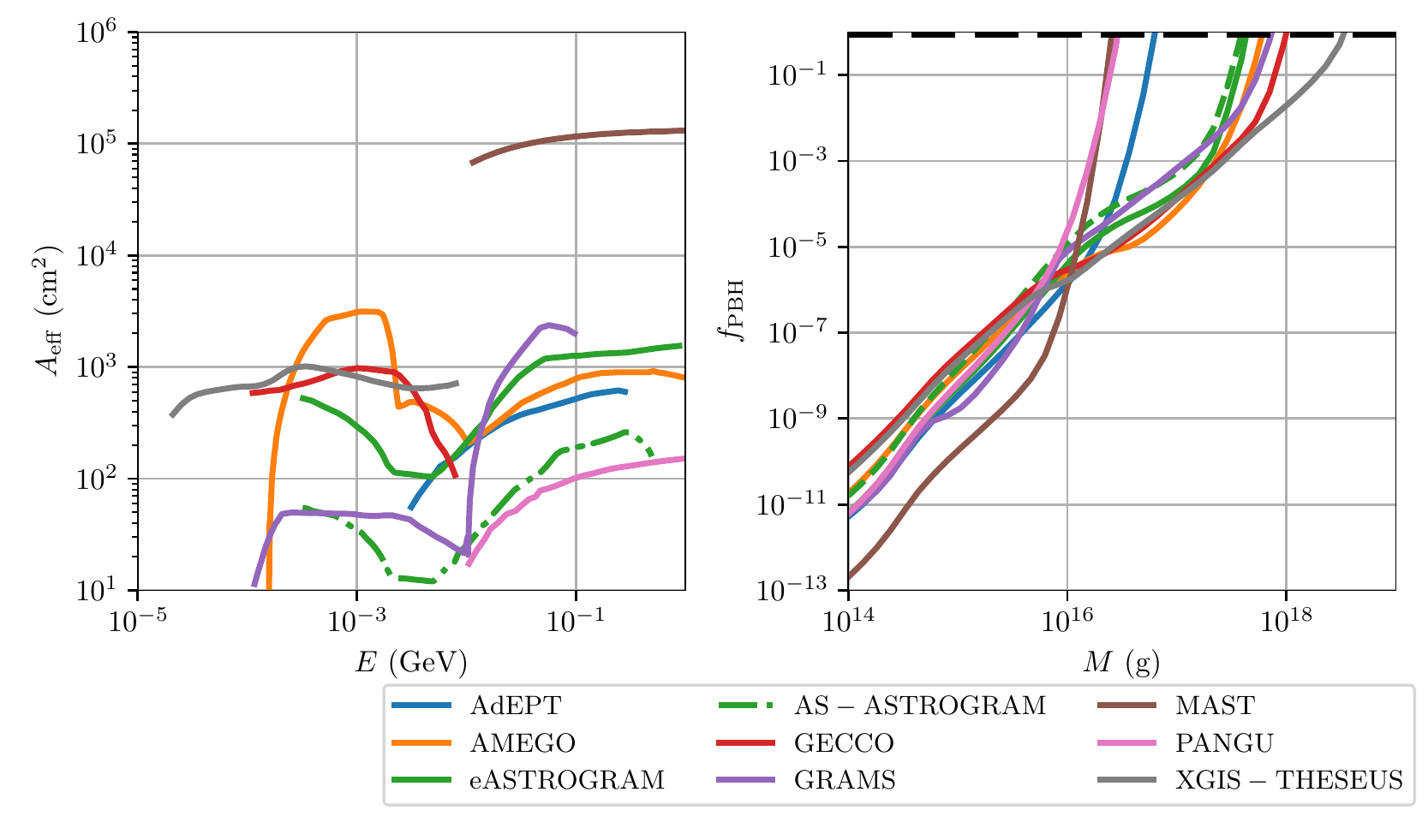}
    \caption{\textbf{Left:} Effective area for prospective optical instruments (see text), adapted from Refs.~\cite{Coogan:2020tuf,Coogan:2021sjs,Ghosh:2021gfa,Dzhatdoev:2019kay}. We have truncated the effective area of MAST as we are not concerned in $>\,$GeV photons for the PBH mass range probed here.
    \textbf{Right:} PBH constraints from GC+EXGB observations using the method M$_2^2$ with the background of~\cite{Bartels:2017dpb}.
    }
    \label{fig:prospective}
\end{figure*}

A large set of prospective instruments, among which AdEPT, AMEGO, ASTROGRAM (results for the AS-ASTROGRAM design are new to this work), GECCO, GRAMS, MAST, PANGU and XGIS-THESEUS, are designed to explore the MeV energy range more accurately than previously done with COMPTEL and EGRET, which will improve both the description of the EXGB and the diffuse GC emission. This energy scale is crucial to set constraints on the fraction of DM $f\mrm{PBH}$ in the disputed mass range $10^{15}-10^{18}\,$g. Data is already available in this energy band, so new instruments will only reduce the error bars. The total number of photons --- background plus PBH signal --- observed by a prospective instrument X is given by~\cite{Coogan:2020tuf,Coogan:2021rez,Ghosh:2021gfa}
\begin{equation}
    N^{\rm X}\mrm{tot} = T\mrm{obs} \Delta\Omega \int_{E\mrm{min}}^{E\mrm{max}} \!\!\!\!A\mrm{eff}^{\rm X}(E)\,\d E \int W(E,E^\prime)\, \dfrac{\d\Phi^{\rm X}\mrm{tot}}{\d E^\prime}\,\d E^\prime\,,\label{eq:nb_particles}
\end{equation}
where $T\mrm{obs}$ is the duration of observation, $\Delta\Omega$ is the solid angle field of view (fov), $A\mrm{eff}$ is the effective area (which defines the energy band probed, see Fig.~\ref{fig:prospective}, left panel) and $W(E,E^\prime)$ is a window function accounting for the finite energy resolution of the instruments. Following the literature, we take this to be a Gaussian~\cite{Bringmann:2008kj}.\footnote{Inside \texttt{Isatis}, we have ensured that the relative energy resolution of the photon spectra matches at least that of the instruments, i.e.~less than a percent.} All the useful references concerning these instruments are listed in Table~\ref{tab:instruments} in Appendix~\ref{app:Isatis}. Following method M$_2$, the PBH constraint is obtained by requiring that the SNR stays below some detection threshold
\begin{equation}
    \dfrac{f\mrm{PBH}N\mrm{PBH}}{\sqrt{N\mrm{bckg}}} \le {\rm SNR}\,,\label{eq:prospective}
\end{equation}
where we have separated the PBH and the background contributions to the total photon count. An alternative method would consist in demanding that this SNR is not attained in \textit{any} energy bin of the background. We have checked with \texttt{Isatis} that this results in a relative change of the constraint $f\mrm{PBH}$ up to a factor of a few.

Most of the difficulty resides in the choice of background. As discussed above, most backgrounds are calibrated to the data and thus automatically contain feedback loops. In Fig.~\ref{fig:prospective} (right panel), we show the PBH constraints obtained by using the same fitted background as Ref.~\cite{Coogan:2020tuf} (see~\cite{Bartels:2017dpb} for details), i.e.~method M$_2^2$; with ${\rm SNR} = 5$, $T\mrm{obs} = 10\,$yr and $\Delta\Omega = 5^o\times 5^o$. The other parameters follow Section~\ref{subsec:existing}. While not reproduced here, we checked that we obtain results very close to that of Ref.~\cite{Coogan:2020tuf} for AdEPT, GECCO and GRAMS (see also \cite{LeyVa:2021kyk}), Ref.~\cite{Ray:2021mxu} for AMEGO and Ref.~\cite{Ghosh:2021gfa} for XGIS-THESEUS. There are discrepancies with Ref.~\cite{Coogan:2021rez} for GECCO and Ref.~\cite{Arbey:2021yke} for AMEGO, probably linked to the different statistical treatment. Stranger are the discrepancies we find with respect to Ref.~\cite{Coogan:2020tuf} for AMEGO, eASTROGRAM, MAST and PANGU: the constraints obtained by~\cite{Coogan:2020tuf} reach up to PBH masses that \textit{cannot} be probed with those 4 instruments; $M\gtrsim 10^{18}\,$g (corresponding to $E\lesssim100\,$keV) for AMEGO and eASTROGRAM, and $M\gtrsim 10^{17}\,$g (corresponding to $E\lesssim 1\,$MeV) for PANGU and MAST. Our constraints look more reasonable to this point of view: only the XGIS-THESEUS instrument can probe $M$ up to several $10^{18}\,$g.

As a second example, we explore the effect of the background choice, that is also a type \textit{iii)} assumption. In Table~\ref{tab:bckg_impact}, we show the extremal relative change in $f\mrm{PBH}$ when choosing the same background as Ref.~\cite{Ghosh:2021gfa} and the agnostic background consisting of the central values of the data points (i.e.~method M$_2^1$), as compared to the background of~\cite{Coogan:2020tuf}. We observe that the choice of background, for the 3 examples shown, can have an impact up to a factor of a few on $f\mrm{PBH}$ in the case of the instrument XGIS-THESEUS.

\begin{table}[t]
    \centering
    \begin{tabular*}{\columnwidth}{l@{\extracolsep{\fill}}c@{\extracolsep{\fill}}c}
        \toprule
        instrument & Ref.~\cite{Ghosh:2021gfa} & data points\\
        \midrule
        AdEPT & $-0.28$ & $-0.08$  \\
        AMEGO & $+0.17$ & $+0.16$ \\
        eASTROGRAM & $+0.11$ & $+0.17$ \\
        AS-ASTROGRAM & $+0.10$ & $+0.16$ \\
        GECCO & $+0.82$ & $+0.72$ \\
        GRAMS & $+0.38$ & $+0.43$ \\
        MAST & $-0.47$ & $-0.02$ \\
        PANGU & $-0.54$ & $-0.02$ \\
        XGIS-THESEUS & $+3.19$ & $+3.54$ \\
        \bottomrule
    \end{tabular*}
    \caption{Relative extremal variation in $f\mrm{PBH}$ when using the same background as~\cite{Ghosh:2021gfa} or the central values of the data points compared to the background of~\cite{Bartels:2017dpb} for the instruments of Fig.~\ref{fig:existing}.}
    \label{tab:bckg_impact}
\end{table}

\section{Reverse engineering the Hawking radiation constraints}
\label{sec:engineering}

In the previous Section, we have revisited existing constraints and exposed their robustness relative to some type \textit{iii)} assumptions, namely the background choice and the statistical method of comparison between the PBH signal and data. In this Section, we examine thoroughly all the other assumptions. Indeed, changing the instrument characteristics at will inside \texttt{Isatis} gives an unprecedented access to a reverse procedure: what would be the capabilities of an ``ideal'' instrument if we could fix arbitrarily all the parameters? Answering this first question carefully could certainly give hints towards design choices for future instruments. But we can go deeper: whatever be the capabilities of the instruments, is there a limit to the PBH mass range we can constrain with \textit{direct} photon PBH constraints? In other words, can we close the window for all DM into PBHs still open between $\sim 10^{17}-10^{22}\,$g? If the answer to this second question is \textit{yes}, this will certainly weigh in the favor of dedicated instruments designs. If it is \textit{no}, theoretical efforts will need to be pursued to use other kinds of constraints in the (to be determined) remaining window.

As already stated, for a given PBH mass $M$, the constraint $f\mrm{PBH}(M)$ for an ``ideal'' prospective instrument is impacted by several assumptions regrouped in 4: \textit{i)} instrument characteristics, \textit{ii)} uncertainties on the (extra)galactic fluxes, \textit{iii)} different statistical treatments and \textit{iv)} uncertainties on Hawking radiation. Each of those can then be decomposed into several contributions, that we review in detail below. We define a test instrument with a given set of technical characteristics, along with fidutial parameters for the (extra)galactic fluxes and standard assumptions for the Hawking radiation, and we fix the statistical method:
\begin{itemize}
    \item[\textit{i)}] observation time $\overline{T}\mrm{obs} = 10\,$yr, fov $\overline{\Delta\Omega} = 5^o\times 5^o$ around the GC, relative energy resolution $\overline{\epsilon}(E) = 1\%$, constant effective area $\overline{A}\mrm{eff}(E) = 10^3\,$cm$^2$ for $E = 1\,{\rm keV}-1\,{\rm GeV}$,
    \item[\textit{ii)}] standard radiation domination from the CMB to today for the extragalactic flux and $\overline{\Omega}\mrm{DM}$ from Planck~\cite{Planck:2018vyg}, NFW profile for the Milky Way halo with the ``convenient'' parameters of~\cite{McMillan:2011wd} resulting in $\overline{J}\mrm{gal}$ (cf.~Section~\ref{subsec:existing}),
    \item[\textit{iii)}] statistical method M$_2^2$ (central values of the data points as background) which considered with point \textit{ii)} gives $\overline{\d\Phi}\mrm{bckg}/\d E$, and $\overline{\rm SNR} = 5$,
    \item[\textit{iv)}] Schwarzschild PBHs with a monochromatic distribution, Hawking primary spectra $\overline{Q}_i$ computed by \texttt{BlackHawk} with the low-energy \texttt{Hazma} branching ratios $\overline{{\rm Br}}_{i\rightarrow j}(E,E^\prime)$ for the secondary spectra.\footnote{In particular, we do assume that the semi-classical Hawking radiation holds on throughout most of the PBH lifetime, see discussion in Sec.~\ref{sec:hawking_rad}.}
\end{itemize}
These characteristics result in a fidutial constraint that we denote by $\overline{f}\mrm{PBH}(M)$. Any modification of some assumption results in a different constraint $f\mrm{PBH}(M)$ that we parametrize through
\begin{equation}
    f\mrm{PBH}(M) = \overline{f}\mrm{PBH}(M) \left( \prod_{\rm \textit{i)},\,\,\textit{ii)},\,\,\textit{iii)},\,\,\textit{iv)}} \alpha({\rm param.}) \right)
\end{equation}
where the $\alpha$'s are the detailed contributions from types \textit{i)-iv)} assumptions. We detail the mathematical form of these in the following, and give quantitative results for the modified $f\mrm{PBH}$ thanks to \texttt{Isatis}.

\subsection{Instruments parameters}

\begin{figure}
    \centering
    \includegraphics[scale = 1]{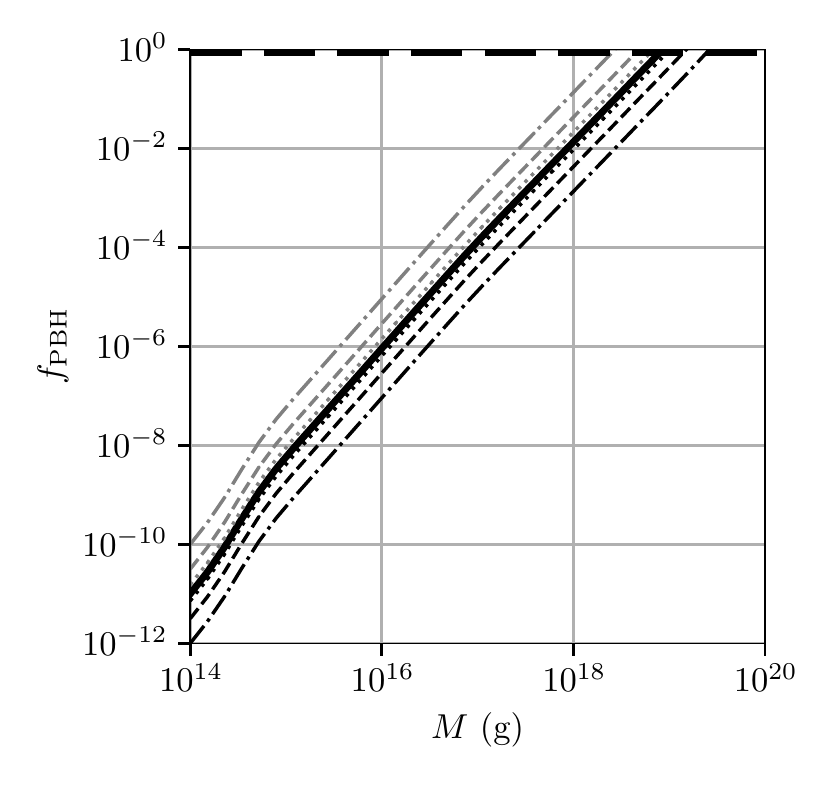}
    \caption{Modification of the constraint $f\mrm{PBH}$ for $T\mrm{obs} = \{1,100\}\,$yr (dashed grey and black), $A\mrm{eff} = \{10^1,10^5\}\,$cm$^2$ (dot-dashed grey and black), $\epsilon = 10\%$ (solid grey, indistinguishable from the fidutial) and $\Delta\Omega = \{2^o\times 2^o,10^o\times 10^o\}$ (dotted grey and black) compared to the fidutial set (solid black). ``Test'' instruments were implemented inside \texttt{Isatis} to obtain these constraints.}
    \label{fig:instruments_parameters}
\end{figure}

A direct look at Eqs.~\eqref{eq:nb_particles} and \eqref{eq:prospective} shows that
\begin{equation}
    \alpha(T\mrm{obs}) = (T\mrm{obs}/\overline{T}\mrm{obs})^{-1/2},\label{eq:alpha_Tobs}
\end{equation}
and
\begin{equation}
    \alpha(A\mrm{eff}) = (A\mrm{eff}/\overline{A}\mrm{eff})^{-1/2},\label{eq:alpha_Aeff}
\end{equation}
as the number of photons captured for both the background and the signal is directly proportional to these quantities. In the limit in which the fov covers only the desired target, we also have
\begin{equation}
    \alpha(\Delta\Omega) \sim (\Delta\Omega/\overline{\Delta\Omega})^{-1/2}.\label{eq:alpha_fov}
\end{equation}
Energy resolutions are overall small and might get smaller in the future as detection techniques improve, so that the window function in Eq.~\eqref{eq:nb_particles} tends towards a Dirac distribution, reducing the pollution from neighbouring energy bins. Thus, changing the energy resolution should have a negligible impact $\alpha(\epsilon) \sim 1$. All these estimations are confronted to quantitative results from \texttt{Isatis} in Fig.~\ref{fig:instruments_parameters}. First, we observe that the fidutial constraint extends up to $\sim 10^{19}\,$g, because the energy coverage of that ``ideal'' instrument goes down to the keV scale. Second, we note that the constraint is close to a power-law of the PBH mass $f\mrm{PBH}(M) \sim (M\,{\rm (g)}/8\times10^{18})^2$, a feature that comes from four facts: the primary emission peaks at an energy that is proportional to the PBH mass; the number density of PBH is inversely proportional to their mass; the background fit of~\cite{Bartels:2017dpb} is approximately a power-law; and the fidutial effective area is a constant. The effect of the secondary spectra is somewhat visible in the slope breaking below $10^{15}\,$g. The global variations of $f\mrm{PBH}$ compared to $\overline{f}\mrm{PBH}$ from the modification of the various parameters has precisely the behaviour expected from Eqs.~\eqref{eq:alpha_Tobs}, \eqref{eq:alpha_Aeff} and \eqref{eq:alpha_fov}. We verify that the energy resolution has no visible impact.

\subsection{Flux parameters}

\begin{figure*}
    \centering
    \includegraphics{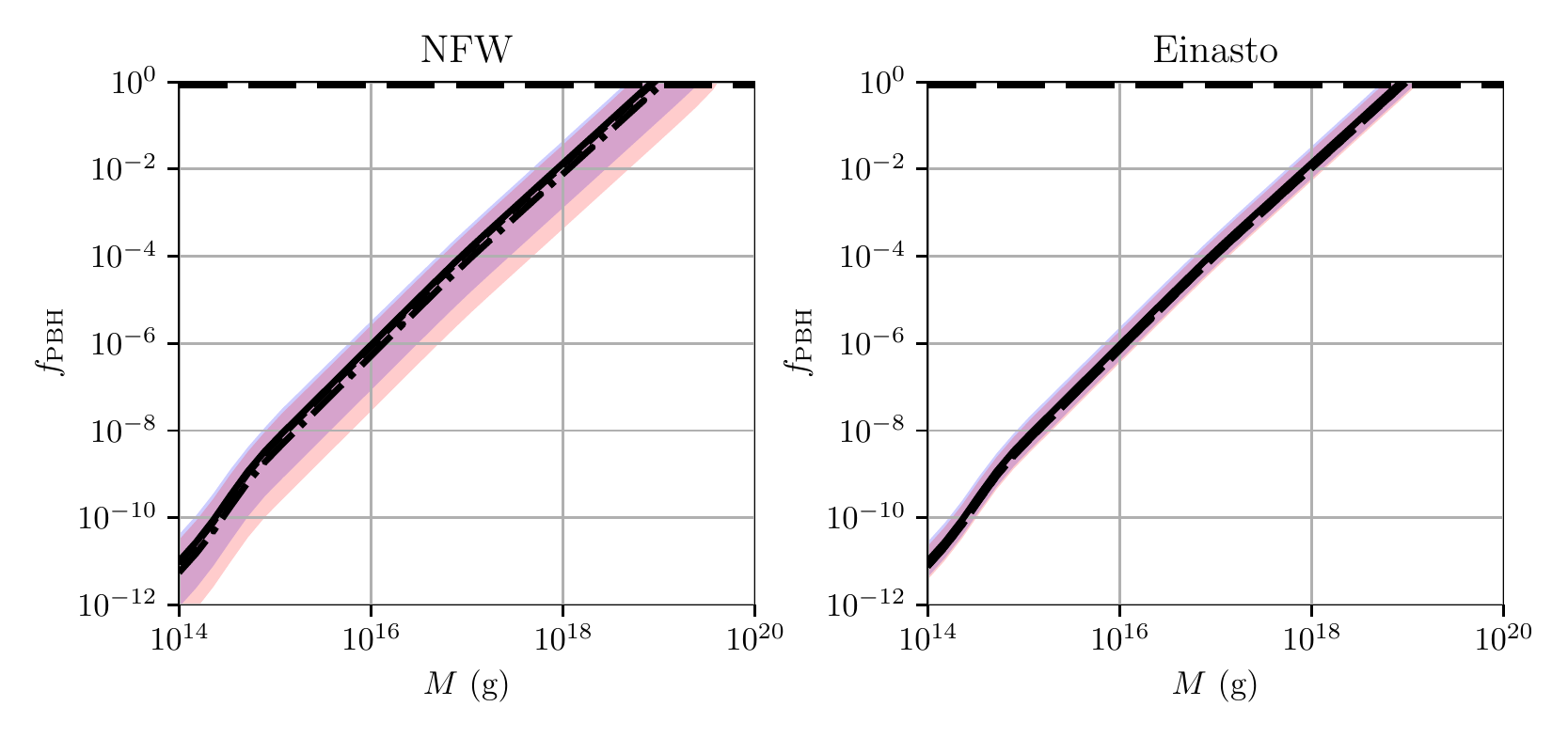}
    \caption{Modification of the constraint $f\mrm{PBH}$ for different galactic profiles, compared to the fidutial case (solid black line).
    \textbf{Left:} $f\mrm{PBH}$ for a generalized NFW profile.
    \textbf{Right:} $f\mrm{PBH}$ for a Einasto profile.
    Dashed (dot-dashed) black lines correspond to the central values of the parameters of~\cite{deSalas:2019pee} and baryonic model B1 (B2); shaded blue (red) areas correspond to a complete span of the $68\%$ CL range for all the parameters for model B1 (B2).}
    \label{fig:profile_impact}
\end{figure*}

As the column density of the target impacts only $N\mrm{PBH}$, we can estimate
\begin{equation}
    \alpha(J\mrm{target}) \sim (J\mrm{target}/\overline{J}\mrm{gal})^{-1},
\end{equation}
The precise impact is obviously more complex depending on the predominance of the target flux over the diffuse extragalactic flux, which also depends on the energy considered. As seen before, we expect the target flux to dominate at high energies, constraining low PBH masses, and the redshifted isotropic flux to dominate at low energies, constraining high PBH masses. In Fig.~\ref{fig:profile_impact} we show the extremal change in $f\mrm{PBH}$ for different galactic DM profiles:
\begin{itemize}
    \item the generalized Navarro--Frenk--White (NFW) profile defined by~\cite{Navarro:1996gj,deSalas:2019pee}
    \begin{equation}
        \rho\mrm{DM}(r) = \rho_c \left( \dfrac{r\mrm{c}}{r} \right)^\gamma\left( 1 + \dfrac{r}{r\mrm{c}} \right)^{\gamma - 3},
    \end{equation}
    where the classical NFW is obtained for $\gamma = 1$
    \item the Einasto profile defined by~\cite{Einasto:1965czb,deSalas:2019pee}
    \begin{equation}
        \rho\mrm{DM}(r) = \rho_c \exp\left\{ -\dfrac{2}{\gamma}\left( \left( \dfrac{r}{r\mrm{c}} \right)^\gamma - 1 \right) \right\}.
    \end{equation}
\end{itemize}
Both of these profiles depend on 3 independent parameters: a characteristic density $\rho\mrm{c}$, a characteristic radius $r\mrm{c}$ and a power $\gamma$. Even with modern measures, these parameters suffer from large uncertainties. We use the $68\%$ CL parameters of~\cite{deSalas:2019pee} (Tables~II and III corresponding to two baryonic models B1 and B2) to be as general as possible in Fig.~\ref{fig:profile_impact}.\footnote{Refs.~\cite{deSalas:2019pee,Coogan:2020tuf} use the incredibly precise radius of the Sun orbit $R_\odot = 8.122\pm0.031\,$kpc recently obtained by the measurement of the orbit of S2~\cite{GRAVITY:2018ofz}, which differs from the older ``convenient'' value of~\cite{McMillan:2011wd}. Ref.~\cite{Coogan:2020tuf} further uses the central values of the NFW profile of Table~III and maximizes the density parameter $J\mrm{gal}$ for the Einasto profile with the $68\%$ CL values of Table~III.} Spanning the whole $68\%$ CL range of the parameters results in constraints $f\mrm{PBH}$ varying by 2 (resp. 1) orders of magnitude inside the same profile and baryonic model for NFW (resp.~Einasto) profile; by $\sim 50\%$ (resp.~$\sim 1\%$) from one baryonic model to the other (B1 or B2) for NFW (resp.~Einasto) profile; and by $20\%$ (resp.~$30\%$) from one profile to the other (NFW or Einasto) for model B1 (resp.~B2). The largest variations are obtained with the generalized NFW profile. We are far from having explored the complete diversity of the DM halo profiles existing in the literature, but we can already predict that $\alpha({\rm gal}) \sim 10^{-1}-10^{1}$. We do not consider targets other than the GC. Even if the density factor is determined more precisely in e.g.~M31 or DM concentrated dSphs like Draco because we observe them as a whole and are not sensitive to their halo profile, it is much smaller than that of the GC and results in less stringent constraints ($10^{-2}$ factor reduction for M31 and Draco~\cite{Coogan:2020tuf}).

A modified redshift history of the universe or DM density may have a small impact on the time stacked isotropic spectrum. We have checked that this is negligible regarding the small uncertainties on the cosmological parameters for the recent universe (after the CMB), resulting in $\alpha({\rm egal}) \sim 1$.

\subsection{Statistical treatment}

As the photon background only affects $N\mrm{bckg}$, we have already concluded in Section~\ref{subsec:prospective} that
\begin{equation}
    \alpha({\rm bckg}) = \left(\dfrac{\d\Phi\mrm{bckg}}{\d E}/\dfrac{\overline{\d\Phi}\mrm{bckg}}{\d E}\right)^{1/2}, 
\end{equation}
within method M$_2$. The different choices of backgrounds for prospective instruments in Table~\ref{tab:bckg_impact} result in $\alpha({\rm bckg}) \sim 0.5-3$ at most. This impact is completely equivalent to that of changing the CL parameter inside method M$_1 ^1$, as shown by Table~\ref{tab:CL_impact}. On the other hand, the SNR is directly proportional to $f\mrm{PBH}$, thus
\begin{equation}
    \alpha({\rm SNR}) = ({\rm SNR}/\overline{\rm SNR})^{1}.
\end{equation}
As this paper was finalized, the author became aware of Ref.~\cite{Chen:2021ngo}, where the constraints from the EXGB are revisited with data from instruments different from those presented in Fig.~\ref{fig:backgrounds}. The contribution from the GC included, as well as Hawking radiated electron-positron annihilation. Ref.~\cite{Chen:2021ngo} takes into account astrophysical contribution to the diffuse photon flux in order to obtain more stringent constraints on the PBH abundance, thus using what we denoted as method M$_1^2$. They obtain constraints more stringent by 2 orders of magnitude in the mass range $10^{15}-10^{17}\,$g.

\subsection{PBH Hawking radiation}

Last but not least, there are uncertainties related to the very computation of the PBH Hawking radiation spectra. These only affect $N\mrm{PBH}$ in Eq.~\eqref{eq:prospective}, thus we predict
\begin{equation}
    \alpha(Q_\gamma) = (Q_\gamma/\overline{Q}_\gamma)^{-1},
\end{equation}
and
\begin{equation}
    \alpha({\rm Br}_{i\rightarrow\gamma}) = \left( {\rm Br}_{i\rightarrow\gamma}/\overline{\rm Br}_{i\rightarrow\gamma} \right)^{-1}.
\end{equation}

The uncertainties on the primary emission rates for Schwarzschild PBHs are directly related to the tables contained inside \texttt{BlackHawk}, which have been confronted to the literature with an accuracy of less than a percent, such that $\alpha(Q_\gamma) \sim 1$.

There could be large uncertainties in the branching ratios computed by the particle physics codes. At very high energy $E\gtrsim{\rm TeV}$ \texttt{BlackHawk} relies on \texttt{HDMSpectra}~\cite{Bauer:2020jay}, in the documentation of which is explained that the precise treatment of the electroweak cascades can alter the branching ratios into photons by several orders of magnitude compared to \texttt{PYTHIA}. This has not yet been confronted to accelerator data. At LHC energies $E\sim{\rm GeV}-{\rm TeV}$, \texttt{BlackHawk} relies on \texttt{PYTHIA}~\cite{Sjostrand:2014zea} or \texttt{HERWIG}~\cite{Bellm:2015jjp} with good correspondence to the data, even with the QCD uncertainties~\cite{Amoroso:2018qga}.\footnote{Updated estimates of the QCD uncertainties will be released soon~\cite{Jueid:2021dlz}.} At low energies $E\lesssim {\rm GeV}$, \texttt{BlackHawk} relies on \texttt{Hazma}~\cite{Coogan:2019qpu}, which is also based on accelerator data, but does not take into account e.g.~the Bremsstrahlung radiation of charged particles specific to PBH radiation~\cite{Page:2007yr} which should dominate at the keV scale. The choice of the QCD scale $\Lambda\mrm{QCD}$ at which pions are emitted as primary particles and of the dynamical rest masses of the quarks and gluon introduces further uncertainties (see~\cite{Calza:2021czr} for a recent discussion). Ref.~\cite{Bauer:2020jay} has shown that the branching ratios for ``low final energies'' $E/E^\prime \lesssim 10^{-4}$ in Eq.~\eqref{eq:secondary} as computed with \texttt{PYTHIA} (and thus \texttt{HDMSpectra}) suffer from order-of-magnitude uncertainties linked to the difficult tracing of electroweak cascades on such stretched scales. Hence, we can conclude that even with very precise primary spectra, the secondary spectra (integrated over primary energies from all scales) are associated with an order-of-magnitude possible variation from all these codes $\alpha({\rm Br}_{i\rightarrow\gamma}) \sim 10^{-1}-10^{1}$.

\subsection{Summary}

\begin{figure}
    \centering
    \includegraphics{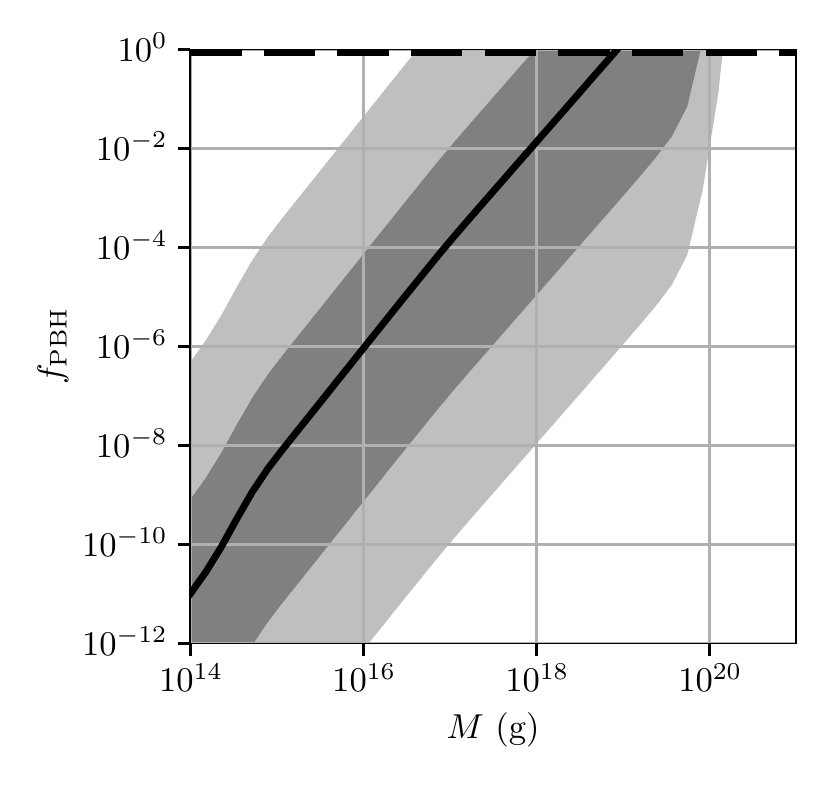}
    \caption{Schematic PBH constraints for all extremal assumptions $\alpha\mrm{tot}$ (light grey area) and with only the extremal fidutial instrument characteristics $\alpha_{\left.i\right)}$ (dark grey area), compared to the fidutial case (dark solid line).}
    \label{fig:extremal}
\end{figure}

Summarizing the results from this Section, we see that compared to our fidutial case $\overline{f}\mrm{PBH}$, the value of the PBH constraint for all masses can vary within each set of assumptions:
\begin{itemize}
    \item[\textit{i)}] $\alpha_{\left.i\right)} = \alpha(T\mrm{obs})\times\alpha(A\mrm{eff})\times\alpha(\Delta\Omega)$, with extremal values $\alpha_{\left.i\right)}^{\rm min} \approx 8\times 10^{-4}$ for $T\mrm{obs} = 100\,$yr, $A\mrm{eff} = 10^5\,$cm$^2$ and $\Delta\Omega = 4\pi$ and $\alpha_{\left.i\right)}^{\rm max} \approx 80$ for $T\mrm{obs} = 1\,$yr, $A\mrm{eff} = 10\,$cm$^2$ and $\Delta\Omega = 2^o\times 2^o$,
    \item[\textit{ii)}] $\alpha_{\left.ii\right)} \approx \alpha({\rm gal})$ with extremal values $\alpha_{\left.ii\right)}^{\rm min} \approx 0.1$ and $\alpha_{\left.ii\right)}^{\rm max} \approx 10$ for the NFW profile,
    \item[\textit{iii)}] $\alpha_{\left.iii\right)} = \alpha({\rm bckg})\times \alpha({\rm SNR})$, with extremal values $\alpha_{\left.iii\right)}^{\rm min} \approx 0.1$ for ${\rm SNR} = 1$ and $\alpha_{\left.iii\right)}^{\rm max} \approx 6$ for ${\rm SNR} = 10$,
    \item[\textit{iv)}] $\alpha_{\left.iv\right)} \approx \alpha(\overline{\rm Br}_{i\rightarrow\gamma})$ with extremal values $\alpha_{\left.iv\right)} \approx 0.1$ and $\alpha_{\left.iv\right)} \approx 10$.
\end{itemize}
Overall, we obtain
\begin{equation}
    8\times10^{-7} \lesssim \alpha\mrm{tot} \lesssim 4.8\times 10^{4} \,,
\end{equation}
with all the sources of uncertainties and
\begin{equation}
    10^{-3} \lesssim \alpha_{\left.ii\right)}\times \alpha_{\left.iii\right)}\times \alpha_{\left.iv\right)} \lesssim 6\times 10^{2} \,,
\end{equation}
if we restrict ourselves to the fidutial instrumental parameters, which is still, as we say in french, \textit{une sacr\'ee fourchette} --- a very large range, as it runs over 11 orders of magnitude for $\alpha\mrm{tot}$ and 5 orders of magnitude for $\alpha_{\left.ii\right)-\left.iv\right)}$. The extremal constraints are schematically shown on Fig.~\ref{fig:extremal}. The $100\%$ PBH-DM scenario is excluded up to $M \sim 10^{20}\,$g in the most favorable case but the open window goes down to $M\sim 10^{17}\,$g in the less favorable one. Of course, the extremal values of the (independent) uncertainties \textit{ii)-iv)} are presumably not attained at the same time, but nevertheless we see that \textit{PBH constraints from direct Hawking radiation of photons are still highly model dependent} due to several sources of uncertainties. We see that modifying the prospective instrument characteristics allows to close the $M \lesssim 10^{20}\,$g window, but we also note that the slope of $f\mrm{PBH}$ is nearly vertical in this region; we discuss this last feature below.

\subsection{Very low energies}

We observe on Fig.~\ref{fig:extremal} that the constraint $f\mrm{PBH}$ saturates at high PBH masses. This is due to the exponential cutoff of $Q\mrm{\gamma}$ and thus of $\d\Phi\mrm{PBH}/\d E$ at energy $E\gtrsim E\mrm{max}\sim T\mrm{PBH} \approx 1\,{\rm keV}(M/10^{19}\,{\rm g})^{-1}$ for Schwarzschild PBHs. Hence, as our description of the background and of the secondary photon spectrum is limited to this keV lower energy bound in this study, we cannot obtain $f\mrm{PBH}$ at $M\gtrsim 10^{20}\,$g. Increasing the capabilities of our prospective ``ideal'' instrument would only make the constraint steeper at $M \sim 10^{20}\,$g. Thus, pushing the PBH constraint on \textit{direct} photon emission up to lower energies and thus higher PBH masses, with the aim of constraining the whole (yet) open window $M \sim 10^{18}-10^{22}\,$g for all DM into PBHs, would require both precise background description at $E \sim 10-1000\,$eV \textit{and} precise particle physics codes to compute electroweak showers down to the eV energy range. We can suppose that the primary spectrum dominates below the keV scale as the emission of all particles except for photons, neutrinos and gravitons is exponentially suppressed at $M\gtrsim 10^{17}\,$g. However, this energy range falls down right in the (very) far UV band ($\lambda \sim 10-1000\,$\AA). To our knowledge, this band is not yet covered within a precise all-sky (resp.~GC) survey that would give an access to the isotropic (resp.~GC) component of the background. The GALEX~\cite{Henry:2014jga} (and future UVEX~\cite{Kulkarni:2021tit}) instruments are sensitive only starting above $\lambda\sim 1000\,$\AA.

Suppose that we extrapolate the ``ideal'' instrument capabilities, the PBH Hawking radiation rates and the GC+EXRB background down to the eV energy scale: $\d\Phi\mrm{bckg}/\d E\propto E^{-2}$ and $A\mrm{eff}(E) = \overline{A}\mrm{eff}(E)$ at $E = 10-1000\,$eV. For the fidutial values of the parameters \textit{i)}$-$\textit{iv)} we obtain $f\mrm{PBH} \sim 10^{6}$ at $M = 10^{22}\,$g. Closing the window for all DM into PBHs with \textit{direct} photons would then require to increase the capabilities of our ``ideal'' instrument by a factor $10^6$ (or reduce $f\mrm{PBH}$ by a factor $10^{-6}$), which is unrealistic as $\alpha_{\left. i\right)}^{\rm min} = 8\times 10^{-4}$ is already technically challenging. Plus, we expect the spectrum of light at these wavelengths to be overcrowded by astronomical sources (see e.g.~Fig.~2 of~\cite{Kulkarni:2021tit}), further increasing the amelioration factor needed to obtain $f\mrm{PBH} \sim 1$ at $M = 10^{22}\,$g. Thus, we conclude that the window for all DM into PBHs cannot be closed by \textit{direct} photon detection from PBH evaporation.\footnote{We note that Ref.~\cite{Mittal:2021dpe} concludes similarly to impossible detection of direct $21\,$cm radio photons at $\mu$eV scale and PBH masses $10^{17}-10^{28}\,$g, and Ref.~\cite{Arbey:2020urq} with direct photons emitted by a $\sim 10\,M_\oplus$ ``Planet 9'' PBH in the Solar System.} Complementary constraints, relying on complex modelling, are thus needed: dynamical capture by or microlensing of stars (see the review by~\cite{Montero-Camacho:2019jte}), rare collisions with solid objects~\cite{Yalinewich:2021fdr} or GRB lensing~\cite{Jung:2019fcs}.

\section{Conclusion}
\label{sec:conclusion}

In this paper, we have computed the photon emission spectrum by Hawking radiation of monochromatic Schwarzschild primordial black holes. We restricted ourselves to the \textit{direct} photons resulting from the primary emission of photons and the secondary emission resulting from electroweak interactions or decay of other primary particles. We obtained the flux on Earth of photons coming from the galactic center plus the isotropic redshifted emission from past ages. We identified all the assumptions underlying the computation of this flux and ended up with a clear nomenclature of the constraining methods M$_i^j$, classifying existing and prospective constraints. Then, we used \texttt{Isatis}, a new public tool available inside the Hawking radiation code \texttt{BlackHawk} and described in the Appendix, to examine quantitatively how the constraints presented in the literature depend on those assumptions, regrouped in 4 categories: \textit{i)} instrument characteristics, \textit{ii)} (extra)galactic flux description, \textit{iii)} statistical method used and \textit{iv)} Hawking radiation uncertainties. We have shown that, for a given instrument design \textit{i)}, the constraint on the DM fraction into PBHs $f\mrm{PBH}(M)$ can span up to 5 orders of magnitude due to uncertainties \textit{ii)}-\textit{iv)}. This has a direct impact on the size of the current and prospective available window for the $100\%$ PBH-DM scenario and questions the robustness of the current constraints. \textit{We underline the necessity of more precise galactic photon background and Hawking radiation determination}, especially at low energies where high mass PBH constraints could close the window. We emphasize that the very validity of the semi-classical computation of Hawking radiation assumed throughout the paper, that could be included as one of the \textit{iv)} uncertainties, is a fundamental basis of the whole work. Finally, we have examined to what extent a prospective ``ideal'' instrument could constrain $f\mrm{PBH} < 1$ from (currently) $M\sim 10^{18}\,$g up to $M\mrm{max}\sim 10^{20}\,$g by ``reverse engineering'' \texttt{Isatis}. We conclude that even if \textit{i)} instrument characteristics allow for another 5 orders of magnitude variation of $f\mrm{PBH}(M)$, \textit{above $M\mrm{max}$, direct photon constraints are not effective anymore so that complementary constraints should be developed instead}.

There are numerous ways to build up on this study and improve \texttt{Isatis}. First, we have only considered Schwarzschild PBHs, but some early matter domination models predict PBHs born with high spin, which would increase their photon yield and result in more stringent constraints. Non-standard BH solutions, e.g.~derived from effective loop quantum polymerization, also predict very different photon spectra. Deviations from the semi-classical Hawking radiation computation at mass scales greater than the Planck scale could also be explored. Second, we have considered a monochromatic mass distribution, which is obviously unrealistic and could be refined within a specific PBH formation model. The main effect is to ``spread'' the constraints towards higher and lower PBH masses due to the high and low mass tails of the distribution under consideration. Third, we have computed the \textit{direct} photon spectrum from instantaneous Hawking radiation, but we could go further and obtain the \textit{indirect} photon spectrum after model dependent interactions of \textit{direct} photons or electrons/positrons with the interstellar medium. One last possibility is to constrain PBHs through their \textit{(in)direct} electron/positron, neutrino, graviton or (putative) DM particle yield. This is left for future work, but a thorough list of references can be found on the \texttt{BlackHawk} website (for those who rely on the code) or in the latest review~\cite{Carr:2021bzv}. Finally, we have assumed that PBHs are not clustered when examining the Galactic constraints, but are rather smoothly distributed following the halo density function. Clustered PBHs would appear as point-like sources and would require a suited search.

\newpage

\appendix

\section{Isatis: a public tool to compute consistent PBH constraints}
\label{app:Isatis}

This Appendix is dedicated to the presentation of \texttt{Isatis}, a new public numerical tool to compute PBH Hawking radiation constraints with a controlled set of assumptions.

\subsection{Technical aspects}

\texttt{Isatis} is a public \texttt{C} tool that relies on the \texttt{BlackHawk}~\cite{Arbey:2019mbc,Arbey:2021mbl} code to compute PBH constraints. Be careful to download the last version of \texttt{BlackHawk} (\texttt{v2.1} as we write this article) before using it. \texttt{Isatis} is available as one of the ``add-ons'' (\texttt{/scripts} folder) of \texttt{BlackHawk} and as such can be downloaded together with the main code on the same website:
\begin{center}
    \url{https://blackhawk.hepforge.org}
\end{center}
The \texttt{Isatis} main folder contains:
\begin{itemize}
    \item the main file \texttt{Isatis\_photons.c} that defines all the \texttt{Isatis} routines,
    \item a parameter file \texttt{parameters.txt} that defines the model parameters for the constraints (galactic and extragalactic fluxes assumptions, statistical comparison method),
    \item a file \texttt{README.txt} that summarizes the basic steps in the use of \texttt{Isatis},
    \item a file \texttt{plotting.py} that can be used to plot the constraints obtained with \texttt{Isatis},
    \item a folder \texttt{/constraints} that contains all kinds of numerical tables used to compute the constraints (e.g.~tabulated background fluxes or cross-sections),
    \item a folder \texttt{/BH\_launcher} that we explain below.
\end{itemize}

\subsection{The BH\_launcher program}

The folder \texttt{/BH\_launcher} contains another \texttt{BlackHawk} add-on: an automatic comprehensive launcher adapted to \texttt{Isatis}. With this program, you can launch several \texttt{BlackHawk} runs in parallel in order to obtain a set of spectra for e.g.~different PBH initial masses. The script is interactive so its use should be transparent, but a \texttt{README.txt} file is provided anyway. First, please ensure that \texttt{BH\_launcher.c} contains the correct path to your \texttt{BlackHawk} version and that the \texttt{BlackHawk} programs \texttt{BlackHawk\_inst} and \texttt{BlackHawk\_tot} are themselves compiled. We advise to compile it with \texttt{\#define HARDTABLES} to save time at execution. To obtain the executable \texttt{BH\_launcher.x} type the command \texttt{make} in the terminal from the \texttt{/BH\_launcher} folder. \texttt{BH\_launcher.x} dos not require any argument so just run it with the command \texttt{./BH\_launcher.x}. The script launches \texttt{BlackHawk} the desired amount of times with the given parameters and generate several outputs:
\begin{itemize}
    \item the \texttt{BlackHawk} output is stored as usual inside the \texttt{BlackHawk/results/} folder,
    \item \texttt{BlackHawk/nohup\_*.txt} files are generated for each run of \texttt{BlackHawk},
    \item a file \texttt{BH\_launcher/*} is generated that contains the number of runs and the name of each run, that may be given as an argument to \texttt{Isatis}.
\end{itemize}
Launching many (tens of) parallel executions of \texttt{BlackHawk} can be memory consuming for both RAM and disk. We advise that you tweak the files \texttt{write\_*.txt} in \texttt{BlackHawk/src/tables} in order to keep only the photon output that is needed by \texttt{Isatis}. With \texttt{BH\_launcher}, you can choose the type \textit{iv)} assumptions, namely you can play with the capabilities of \texttt{BlackHawk} concerning non-standard PBHs, extended mass distributions, etc.

\subsection{The Isatis program}

The \texttt{Isatis} main program takes two arguments: a parameter file that contains all the assumptions of types \textit{ii)-iii)} to use in the constraint determination and a file which contains the number of \texttt{BlackHawk} runs \texttt{nb\_runs = *} that you want to compute PBH constraints from, and the names of those runs as a list. The latter can either be generated automatically by the side program \texttt{BH\_launcher} or by hand, listing the results folders of interest that the user may already have generated with \texttt{BlackHawk}. The constraints computation requires that the secondary spectra of particles are computed, and that both programs \texttt{BlackHawk\_inst} and \texttt{BlackHawk\_tot} have been used on the same parameters file. Otherwise, errors are cast when trying to read non-existing output files. Please ensure that \texttt{parameters.txt} contains the correct path to your version of \texttt{BlackHawk}. To compile \texttt{Isatis.c} into \texttt{Isatis.x} just go into the \texttt{/Isatis} main folder and type \texttt{make Isatis}. Then, launch it  with the parameter file \texttt{*} and the run list file \texttt{**} as an argument with the command \texttt{./Isatis.x * **}. This should generate a result file \texttt{results\_photons\_*.txt} containing the list of PBH constraints with each line being one of your \texttt{BlackHawk} runs (e.g.~with some specific PBH initial mass), and each column the corresponding PBH constraint set by a specific (existing or prospective) optical instrument.

To summarize, a basic use of the \texttt{Isatis} program consists in (supposing everything has been compiled correctly):
\begin{enumerate}
    \item launching \texttt{BH\_launcher.x} to generate a set of \texttt{BlackHawk} spectra with some varying PBH parameters,
    \item launching \texttt{Isatis.x} to compute the corresponding PBH photon constraints,
    \item using \texttt{plotting.py} to visualize the constraints.
\end{enumerate}

\subsection{Instrument implementation}

For existing instruments, we use the most recent data they have produced. Those data are presented in Fig.~\ref{fig:backgrounds} together with the corresponding literature. The measured photon fluxes in the direction of the desired target (galactic center, isotropic flux) are tabulated in files \texttt{constraints/photons/flux\_*\_**.txt}. These are used to obtain the constraints labelled M$_1$ in the nomenclature of Fig.~\ref{fig:tree}. For prospective instruments, instrument characteristics (type \textit{i)} assumptions) are directly hardcoded inside \texttt{Isatis}, with relevant literature listed in Table~\ref{tab:instruments}:
\begin{itemize}
    \item the time of observation $T\mrm{obs}$,
    \item the fov $\Delta\Omega$,
    \item the effective area as a function of photon energy $A\mrm{eff}(E)$ (tabulated in files \texttt{constraints/photons/Aeff\_*\_**.txt} where \texttt{*} is the instrument name and \texttt{**} the arXiv identifier),
    \item the relative energy resolution as a function of photon energy $\epsilon(E)$.
\end{itemize}
We propose several (extra)galactic backgrounds inside \texttt{Isatis}, labeled as \texttt{constraints/photons/background\_**.txt}. These are used to obtain the constraints labelled M$_2$ in Fig.~\ref{fig:tree}. All files contain \texttt{*} the name of the instrument and/or \texttt{**} the arXiv identifier to help traceability.

\begin{table}[t]
    \centering
    \begin{tabular*}{\columnwidth}{l@{\extracolsep{\fill}}c@{\extracolsep{\fill}}c}
        \toprule
        instrument & $A\mrm{eff}$ & $\epsilon$ \\
        \midrule
        AdEPT~\cite{Hunter:2013wla} & \cite{Coogan:2020tuf,Coogan:2021sjs} & \cite{Hunter:2013wla} \\
        AMEGO~\cite{AMEGO:2019gny,Kierans:2020otl}\footnote{See also \url{https://asd.gsfc.nasa.gov/amego/files/AMEGO_Decadal_RFI.pdf}.} & \cite{Kierans:2020otl,Coogan:2020tuf,Coogan:2021sjs} & \cite{AMEGO:2019gny,Kierans:2020otl} \\
        eASTROGRAM~\cite{e-ASTROGAM:2016bph,e-ASTROGAM:2017pxr} & \cite{e-ASTROGAM:2016bph,Coogan:2020tuf} & \cite{e-ASTROGAM:2016bph,e-ASTROGAM:2017pxr} \\
        AS-ASTROGRAM~\cite{Mallamaci:20198W} & \cite{Coogan:2021sjs} & \cite{Mallamaci:20198W} \\
        GECCO~\cite{Orlando:2021get}\footnote{See also \url{https://www.snowmass21.org/docs/files/summaries/CF/SNOWMASS21-CF3_CF1_Stefano_Profumo-007.pdf}.} & \cite{Coogan:2020tuf,Coogan:2021sjs} & \cite{Coogan:2021rez,Orlando:2021get} \\
        GRAMS~\cite{Aramaki:2019bpi,Aramaki:2020gqm,GRAMS:2021tax} & \cite{Coogan:2020tuf,Coogan:2021sjs,LeyVa:2021kyk,GRAMS:2021tax} & \cite{Aramaki:2019bpi} \\
        MAST~\cite{Dzhatdoev:2019kay} & \cite{Dzhatdoev:2019kay} & \cite{Dzhatdoev:2019kay} \\
        PANGU~\cite{Wu:2014tya,Wu:2016ky} & \cite{Wu:2016ky,Coogan:2020tuf,Coogan:2021sjs} & \cite{Wu:2014tya,Wu:2016ky} \\
        XGIS-THESEUS~\cite{Labanti:2021gji,Campana:2021ith} & \cite{Labanti:2021gji,Campana:2021ith,Ghosh:2021gfa} & \cite{Labanti:2021gji,Ghosh:2021gfa} \\
        \bottomrule
    \end{tabular*}
    \caption{Prospective optical instruments references for effective area $A\mrm{eff}(E)$ (see also Fig.~\ref{fig:prospective}, left panel) and energy resolution $\epsilon(E)$, as implemented into \texttt{Isatis}. All instruments share $T\mrm{obs} = 10\,{\rm yr}\,\approx 10^{8}\,$s.}
    \label{tab:instruments}
\end{table}

\subsection{For information: what's in the literature?}

\begin{table}[t!]
    \centering
    \begin{tabular*}{\columnwidth}{l@{\extracolsep{\fill}}c@{\extracolsep{\fill}}c}
        \toprule
        instrument & target & statistical method \\
        \midrule
        joint isotropic~\cite{Ballesteros:2019exr} & EXGB & M$_1^1$, M$_1^2$ \\
        joint isotropic~\cite{Arbey:2019vqx} & EXGB & M$_1^1$ \\
        joint isotropic~\cite{Carr:2020gox,Carr:2020xqk,Carr:2021bzv} & EXGB & M$_1^2$ \\
        INTEGRAL~\cite{Laha:2020ivk} & GC & M$_1^1$, M$_1^2$ \\
        COMPTEL~\cite{Coogan:2020tuf} & GC & M$_1^1$ \\
        \bottomrule
    \end{tabular*}
    \caption{Existing constraints classification. Joint isotropic: all or part among HEAO + balloon + SMM + COMPTEL + EGRET + Fermi-LAT instruments (ordered with increasing energy domains). EXGB: extragalactic gamma/X-ray background. GC: galactic center.}
    \label{tab:existing_constraints}
\end{table}

\begin{table}[t!]
    \centering
    \begin{tabular*}{\columnwidth}{l@{\extracolsep{\fill}}c@{\extracolsep{\fill}}c}
        \toprule
        instrument & target & statistical method \\
        \midrule
        MeV panel~\cite{Coogan:2020tuf} & GC, M31, Draco & M$_2^2$ \\
        XGIS-THESEUS~\cite{Ghosh:2021gfa} & GC + EXGB & M$_2^1$, M$_2^1$ \\
        GECCO~\cite{Coogan:2021rez} & GC, M31, Draco & M$_2^2$ \\
        AMEGO~\cite{Arbey:2021yke} & GC + EXGB & M$_2^3$ \\
        \bottomrule
    \end{tabular*}
    \caption{Prospective constraints classification. MeV panel: AdEPT, AMEGO, eASTROGRAM, GECCO, GRAMS, MAST PANGU (alphabetical order), taken separately. ``+ EXGB'' indicates that the target mentioned has been considered on top of the EXGB contribution.}
    \label{tab:prospective_constraints}
\end{table}

For transparency, we classify the PBH \textit{direct} photons constraints from existing (Table~\ref{tab:existing_constraints}) and prospective (Table~\ref{tab:prospective_constraints}) instruments that are present in the literature and listed in the Introduction. This is not an exhaustive list as it is restricted to the bounds usually shown in the recent review efforts. This classification could obviously be extended to older or to be published constraints. In this study, we have used method M$_1^1$ for existing instruments and compared methods M$_2^1$ and M$_2^2$ for prospective instruments.

\subsection{Perspectives}

Modification of \texttt{Isatis} should be straightforward. Adding a new instrument or a new background consists in providing the relevant \texttt{*.txt} flux or effective area file. New galactic profiles can be easily implemented, as well as targets different from the GC, thanks to the corresponding density factor $J$. The code is already adapted to extended mass functions with general expressions of the normalization constants $A$ of Eqs.~\eqref{eq:flux_gal} and \eqref{eq:flux_egal}. \texttt{BH\_launcher} is also handy to modify, as it works following a decision tree. Variation of parameters different from the mass from one run to the next consists in adding branches to that tree.

In the near future, we intend to produce neutrino and electron-positron tools as well, in the form of additional programs \texttt{Isatis\_neutrinos.c} and \texttt{Isatis\_electrons.c}. The scheme will be the same as for \texttt{Isatis\_photons.c} but adapted to constraints set for \textit{direct} neutrinos and electrons-positrons. \textit{Indirect} detection constraints are also a horizon of development.

\newpage

\bibliography{biblio}

\end{document}